\pdfoutput=1

\documentclass[]{itmo-student-thesis}

\usepackage{latexsym}
\usepackage{url}
\usepackage{graphicx}
\usepackage{caption}
\usepackage{subcaption}

\usepackage{verbatim}
\usepackage{tikz}
\usetikzlibrary{arrows}

\begin{document}

\studygroup{M3439}
\title{Construction and Quality Evaluation of \\ Heterogeneous Hierarchical Topic Models}
\author{Belyy Anton}{Belyy A.}
\supervisor{Filchenkov Andrey, Ph.D.}{Filchenkov A.}{Ph.D.}{}
\publishyear{2018}
\startdate{01}{sept}{2017}
\finishdate{31}{may}{2018}

\addconsultant{KV}{KV}

\secretary{ON}

\technicalspec{}
\plannedcontents{}
\plannedsources{}

\researchaim{}
\researchtargets{}
\advancedtechnologyusage{}
\researchsummary{}
\researchfunding{}

\researchpublications{}

\maketitle{Bachelor}

\tableofcontents

\startprefacepage
Structuring knowledge and finding relevant literature have always been important problems in science and education. Library classifications \cite{chan2015cataloging}, used to facilitate search of books in the libraries, have been known since 5th century BC, although the first modern ones date back to 19th century AD. 
In terms of functionality, these classifications ranged from simple \textit{enumerative} ones, where each book was assigned to a particular subject and subjects were listed alphabetically, to \textit{hierarchical}, in which subjects were additionally divided from most generic to most specific, and \textit{faceted}, where each book could be assigned to multiple mutually exclusive subjects.

However, library classifications cannot be applied efficiently for structuring and retrieving of large amount of data available on the Internet. Firstly, these classifications were optimised for providing an ability to quickly locate a book on a shelf. It simplified the retrieval of books for library staff, but current on-line storage facilities can hardly benefit from this. Secondly, the catalogs in libraries have to be maintained by hand: the addition of new topics or deletion of old ones, as well as assigning each book to a topic, is done by human experts. This is infeasible for billions of documents available in the World Wide Web. Thirdly, we have to account for diverse (heterogeneous) structure of documents in the Web. While libraries mostly contain books and scientific journals, the Internet additionally provides article pre-prints, tutorials, lecture notes, code snippets, cooking recipes, opinion pieces, news streams, social media responses, and much more.

To provide a tool for learning and investigating documents from various Internet sources, \textit{exploratory search engines (ESEs)} \cite{white2009exploratory, Marchionini2006} have been proposed. Unlike traditional search engines used for looking up previously known and short keywords to answer specific questions, exploratory search systems allow for inexact and arbitrary long queries to explore a broad subject. Much like libraries in the pre-Internet era, these systems attract scholars and students, journalists and analysts, who are interested to find as much information as possible on a given subject up to some level of detail.

Given the difference between traditional and exploratory search engines, one could expect that methods and models used for former systems cannot be directly applied for the latter. We argue that in order to build a search index for an ESE, one should consider at least two key aspects of such engines: 1) ``topical'' nature of both queries and responses, 2) hierarchical nature of topics. We will discuss this aspects in more detail in the next section. To meet the first criterion, we use additive regularization for topic models (ARTM) approach \cite{vorontsov2015additive}. The authors of \cite{ianina2017multi}have shown that this approach performs very well for topic exploration tasks, matching and surpassing human experts' performance. To meet the second criterion, we use hierarchical extension of ARTM \cite{Chirkova2016}.

Probabilistic topic models of textual collections allow to represent a concept of a topic as a conditional probability distribution over words, and to represent a single document as a distribution over topics. For larger collections topics are divided into more detailed \textit{sub-topics}. \textit{Topical hierarchies} in hierarchical topic models (HTMs) are conditional probability distributions representing ``soft'' clusterization of topics into sub-topics.

Various approaches to evaluate flat topic models have been proposed \cite{bouma2009normalized, newman2010automatic, Mimno2011, Fang2016, Nikolenko2016, nikolenko2017topic}. However, there is still no agreement on the common evaluation framework for HTMs~\cite{Zavitsanos2011}.

In our work, we propose to represent HTM as a set of flat models, or \textit{layers}, and a set of topical hierarchies, or \textit{edges}. We suggest several quality measures for edges of hierarchical models, resembling those proposed for flat models. We conduct an assessment experimentation and show strong correlation between the proposed measures and human judgement on topical edge quality. We also introduce \texttt{heterogeneous} algorithm to build hierarchical topic models for heterogeneous data sources. We show how making certain adjustments to learning process helps to retain original structure of customized models while allowing for slight coherent modifications for new documents. We evaluate this approach using the proposed measures and show that the proposed \texttt{heterogeneous} algorithm significantly outperforms the baseline \texttt{concat} approach. Finally, we implement our own ESE called \texttt{Rysearch}, which demonstrates the potential of ARTM approach for visualizing large heterogeneous document collections.

Our work is organized as follows. In Chapter 1, we formulate the problem in terms of specification for a new ESE, which we implemented, and present an overview of existing exploratory search systems based on topic models. We also formally introduce probabilistic topic modeling, ARTM and hARTM approaches in more detail. Finally, we present existing quality measures for flat topic models.

In Chapter 2, we propose new extrinsic and intrinsic similarity based measures for evaluating topical edges of hierarchical topic models, along with several approaches to aggregate them into a single quality measure of topic hierarchy. We continue by introducing \texttt{concat} and \texttt{heterogeneous} algorithms and describing in detail the characteristics and use case scenarios of the new ESE.


In Chapter 3, we conduct thorough experimentation on heterogeneous Russian popular scientific sources. We describe the details and results of assessment experiment there. We provide a comparison of \texttt{concat} and \texttt{heterogeneous} algorithms. Finally, we demonstrate an empirical approach for automated quality improvement of already built topic models basing on the proposed measures.

\startrelatedwork

\chapter{Literature review}

\section{Problem statement}
We are given a set $\mathcal{D} = (D_1, D_2, \textellipsis, D_n)$ of document collections. Each collection $D_i$ contains $|D_i|$ documents, comprised of several modalities (unigrams, or \textit{words}, and user-annotated \textit{tags}). To omit specifying modality, we will often use the term \textit{token}, which means: a word or a tag.  A collection $D_i$ possesses its own vocabulary $W_i$, which is comprised of tokens of two modalities: $W_i = (W_i^{word}, W_i^{tag})$. A \textit{common vocabulary} $\mathcal{W}$ denotes all the words or tags that occur in some collection-specific vocabulary $W_i$ and is defined as $\mathcal{W} = \Big(\bigcup\limits_{i=1}^{n} W_i^{word}\Big) \cup \Big( \bigcup\limits_{j=1}^{n} W_j^{tag} \Big)$. We define size of vocabulary as a sum of sizes of the vocabulary for each modality: $|W_i| = |W_i^{text}| + |W_i^{tag}|$. A document $d_{ij} \in D_i$ is represented as a set of tokens in each modality, with order of tokens being insignificant (bag-of-words model): $d_{ij} = (d_{ij}^{word}, d_{ij}^{tag})$, where $d_{ij}^{MOD} = (w_{ij1}^{MOD}, w_{ij2}^{MOD}, \textellipsis)$, $ w_{ijk}^{MOD} \in W_i^{MOD}$, and  $MOD \in \{word, tag\}$.

A set of document collections $\mathcal{D}$ is said to be \textit{heterogeneous}, if all of the following is true:
\begin{enumerate}
	\item document collections vary \textit{significantly} in document size: \\ $\min\limits_{i}{|D_i|} \ll \max\limits_{j}{|D_j|}$,
	\item document collections vary \textit{significantly} in vocabulary size: \\ $\min\limits_{i}{|W_i|} \ll \max\limits_{j}{|W_j|}$,
	\item document collections have \textit{significantly} different topical structures.
\end{enumerate}

In each of the above-mentioned cases, the significance of a variation is defined empirically by human experts and we will not formalize it further. ``Topical structure'' in the third criterion is also a semi-empirical term here, which will be formalized in the following sections.

In this setup, we are asked to \textbf{propose an algorithm} for building hierarchical topic models (HTMs) over heterogeneous sources, that will ``respect'' difference of each individual source and produce an overall coherent and easy-to-navigate model. Since these properties are defined informally, we are further asked to \textbf{propose evaluation criteria for HTM} and compare the proposed algorithm to baseline method for aggregating heterogeneous sources into a HTM.

To demonstrate the validity of the proposed algorithm  and quality measures for creating exploratory search systems, we are further asked to \textbf{implement a simple ESE} for two heterogeneous sources, \texttt{Postnauka}\footnote{\url{https://postnauka.ru/about}} and \texttt{Habrahabr}\footnote{\url{https://habr.com/info/about/}}, with navigation, searching and automatic tagging facilities. This should also serve as a demonstration of applicability of hARTM approach (discussed later in more detail) for building hierarchical navigators over large collections, continuing the work of Chirkova et al \cite{Chirkova2016}, but extending it to a more realistic heterogeneous case.

Therefore, the contribution of our work is threefold:
\begin{enumerate}
	\item We propose new evaluation measures for hierarchical topic models,
	\item We propose an algorithm to build hierarchical models in the case of heterogeneous data sources,
	\item We implement an exploratory search engine based on the proposed algorithm.
\end{enumerate}

\section{Probabilistic topic modeling}
In probabilistic topic modeling a document collection is a set of triples $(d, w, t)$ sampled from a discrete distribution $p(d, w, t)$ over $D \times W \times T$. $D$ is a set of documents, $W$ is a collection vocabulary i.e. a set of words (tokens) appearing in documents and $T$ is a set of topics. $d \in D$ and $w \in W$ are observable variables, while $t \in T$ is a latent variable. Thus, a topic model of $D$ consists of probability distributions $p(w|t)$ for each $t \in T$ and $p(t|d)$ for each $d \in D$, while distributions $p(w|d)$ are given. An estimator for the given variables is $p(w|d) = \dfrac {n_{dw}}{n_d}$, where $n_{dw}$ is a number of times the token $w$ appears in the document $d$, $n_d = \sum_{w \in W}n_{dw} $ is a number of words in $d$.

Let us make several standard assumptions. First, the order of tokens $w \in W$ in a document $d \in D$ is not important (bag-of-words model). Second, the tokens distributions over topics  are not document dependent, i.e. $p(w|t, d) = p(w|t)$.

Under the listed assumptions a flat (one-level) topic model is described with the Bayes formula	
$$p(w|d) = \sum_{t \in T}p(w|t)p(t|d), \ d\in D, w \in W.$$

The problem can equivalently be stated as a matrix factorization:
\begin{equation}\label{artm:factor}
F \approx \Phi \Theta,
\end{equation}
or as an optimization task:
\begin{gather*}
\sum\limits_{d \in D} \sum\limits_{w \in W} n_{dw} \ln{\sum\limits_{t \in T} \phi_{wt}\theta_{td}} \rightarrow 
\max\limits_{\Phi, \Theta} \\
 \text{w.r.t.}  \ \phi_{wt} \geq 0, \ \theta_{td} \geq 0,\ \sum\limits_{w \in W} \phi_{wt} = 1, \ \sum\limits_{t \in T} \theta_{td} = 1,
\end{gather*}
where 
\begin{equation}
\label{def:F}
F = [p(w|d)]_{W \times D}
\end{equation}
\begin{equation}
\label{def:Phi}
\Phi = [p(w|t)]_{W \times T}
\end{equation}
\begin{equation}
\label{def:Theta}
\Theta = [p(t|d)]_{T \times D}.
\end{equation}

Let us see that the problem \ref{artm:factor} is not well-posed as it admits infinitely many solutions: for any matrix $S$ of rank $|T|$ we have $\Phi \Theta = (\Phi S)(S^{-1} \Theta) = \Phi' \Theta'$. Typically this is solved by reducing the solution space or by introducing \textit{regularization term} $R(\Phi, \Theta)$ to objective function:

\begin{gather*}
\sum\limits_{d \in D} \sum\limits_{w \in W} n_{dw} \ln{\sum\limits_{t \in T} \phi_{wt}\theta_{td}} + R(\Phi, \Theta) \rightarrow 
\max\limits_{\Phi, \Theta} \\
\text{w.r.t.}  \ \phi_{wt} \geq 0, \ \theta_{td} \geq 0,\ \sum\limits_{w \in W} \phi_{wt} = 1, \ \sum\limits_{t \in T} \theta_{td} = 1.
\end{gather*}

 Actually, we can classify topic modeling approaches by their regularization term. Probabilistic Latent Semantic Analysis (PLSA) \cite{hofmann1999probabilistic} is the basic approach which does not use any regularization:
\begin{equation*}
R(\Phi, \Theta)	= 0
\end{equation*}
Latent Dirichlet Allocation (LDA) \cite{blei2003latent} is another common approach, where it is assumed that columns of $\Theta$ and $\Phi$ matrices are random variables with  distributions $Dir(\alpha), \alpha \in \mathbb{R}^{|T|}$ and $Dir(\beta), \beta \in \mathbb{R}^{|W|}$, respectively. Using Bayes inference we can show it is equivalent to the following regularizer:
\begin{equation*}
R(\Phi, \Theta)	= \sum\limits_{t \in T}\sum\limits_{w \in W}(\beta_w - 1) \ln \phi_{wt} + \sum\limits_{d \in D}\sum\limits_{t \in T}(\alpha_t - 1) \ln \theta_{td}
\end{equation*}
Finally, Additive Regularizations of Topic Models (ARTM) \cite{vorontsov2015additive} approach allows to combine arbitrary set of differentiable functions $R_i(\Phi, \Theta), i=1,...n$ with weights $\tau_i$ into a single additive regularizer:
\begin{equation*}
R(\Phi, \Theta)	= \sum\limits_{i=1}^n\tau_i R_i(\Phi, \Theta)
\end{equation*}
These additive regularizers can encode some apriori knowledge about the topics, which we want to enforce (e.g. topics need to sparse, decorrelated, and coherent), but they can also enforce structures, such as hiearchical relationship between topics. In the following section we will discuss it in more detail.

\section{Hierarchical topic modeling}
In this section, we describe the development of hierarchical topic models. Most of the advancements in this field were designed as extensions for LDA framework, starting with hLDA proposed in 2004 by David Blei, so we will describe these extensions in detail. We will also cover hierarchical ARTM (hARTM) \cite{Chirkova2016}, which is an extension for ARTM framework.
\subsection*{Extensions of LDA framework}
Several works have been proposed from 2004 to 2016 to support hierarchies in state-of-the-art LDA model.

Some of them focus on extending generative process of LDA. In hLDA \cite{Blei2004} topical hierarchy is represented as a tree, where each sub-topic has exactly one parent. hPAM \cite{Mimno2007} overcomes this limitation and represents hierarchies as multilevel acyclic graphs. hHDP \cite{Zavitsanos2011} also use multilevel graph model for hierarchies and additionally provide ways to esimate number of levels and number of topics on each level of hierarchy.

Other works focus on scalability and performance on huge-scale datasets. \cite{Wang2014} proposes a method that is both scalable and interpretable by humans. \cite{pujara2012large} proposes an simple iterative meta-algorithm that builds hierarchies in a top-down fashion using MapReduce. It also provides an open-source implementation, which is an extension of MapReduce LDA \cite{zhai2012mr} package.

\subsection*{Hierarchical ARTM}
Let us define a topic hierarchy as an oriented multiparticle (multilevel) acyclic graph where each level is a flat topic model. The edges connect topics from neighboring levels and represent parent-child relations in the hierarchy.

Let us assume that $l \geq 1$ levels of a topic hierarchy are built, the $l$-th level has the parameters $\Phi^l, \Theta^l$, and $A$ is a set of $l$-th level topics. To build the $(l+1)$-th level, we model $l$-th level distributions of tokens over topics $p(w|a), \ a\in A$ as mixtures of the next level distributions $p(w|t), \ t\in T$, where $T$ is a set of $(l+1)$-th level topics. This leads to a Bayes formula $p(w|a) =  \sum_{t \in T}p(w|t)p(t|a), \ a\in A, w \in W.$

The equivalent matrix factorization problem is
$\Phi^{l+1} \approx \Phi^l \Psi,$
where 
\begin{equation}
\label{def:Psi}
\Psi = [p(t|a)]_{T \times A}
\end{equation}
contains the distributions of the $(l+1)$-th level topics over the $l$-th level topics.

\section{Quality of flat topic models}\label{review:qual}
We briefly survey quality measures that have been proposed for flat topic models. We denote $W^{(t)} \subseteq \mathcal{W}$ as a sequence of $n$ top tokens of a topic $t$, that is an ordered sequence of tokens from $t$-th column of $\Phi$ matrix with highest probability:
\begin{equation*}
	W^{(t)} = (\tilde{w_1}^{(t)}, \tilde{w_2}^{(t)}, \textellipsis, \tilde{w_n}^{(t)}), \phi_{\tilde{w_{1}}t} \geqslant \phi_{\tilde{w_{2}}t} \geqslant \textellipsis \phi_{\tilde{w_{n}}t}.
\end{equation*}
Usually, $n$ is fixed to a small number beforehand. Then, all of the major evaluation measures for flat topic models can be expressed in the following form:
\begin{equation}\label{review:qual_eq}
	\mathrm{Quality}(t) = \frac{1}{C} \sum\limits_{i=1}^{n}\sum\limits_{j=1}^{n} f(\tilde{w_i}^{(t)}, \tilde{w_j}^{(t)}),
\end{equation}
where inner term $f(\tilde{w_i}^{(t)}, \tilde{w_j}^{(t)})$ is a measure of cooccurrence of top tokens $\tilde{w_i}^{(t)} \in W^{(t)}$ and $\tilde{w_j}^{(t)} \in W^{(t)}$, and $C$ is measure-specific norm.

We begin with \textit{coherence} proposed in \cite{Mimno2011}. For a topic $t$ it is defined as
\begin{equation} \label{m:1}
	\mathrm{Coh}(t) = \frac{2}{n(n - 1)} \sum\limits_{i=1}^{n-1}\sum\limits_{j=i+1}^{n} \log \frac{d(\tilde{w_i}^{(t)}, \tilde{w_j}^{(t)}) + \varepsilon}{d(\tilde{w_i}^{(t)})},
\end{equation}
where $d(\tilde{w_i}^{(t)})$ is the \textit{document frequency} of top token $\tilde{w_i}^{(t)}$ (i.e., the number of documents with at least one token $\tilde{w_i}^{(t)}$) and $d(\tilde{w_i}^{(t)}, \tilde{w_j}^{(t)})$ is the \textit{co-document frequency} of top tokens $\tilde{w_i}^{(t)}$ and $\tilde{w_j}^{(t)}$ (i.e., the number of documents containing at least one token $\tilde{w_i}^{(t)}$ and at least one token $\tilde{w_j}^{(t)}$).
Two modifications have proposed to the original coherence. First one is \textit{\mbox{tf-idf} coherence}, which is defined as
\begin{equation} \label{m:2}
	\mathrm{Coh_{tfidf}}(t) = \frac{1}{n(n - 1)} \sum\limits_{\tilde{w_i}^{(t)} \neq \tilde{w_j}^{(t)}} \log \frac{\sum_{d: \tilde{w_i}^{(t)}, \tilde{w_j}^{(t)} \in d} \mathrm{tfidf}(\tilde{w_i}^{(t)}, d) \mathrm{tfidf}(\tilde{w_j}^{(t)}, d) + \varepsilon}{\sum_{d: \tilde{w_i}^{(t)} \in d}\mathrm{tfidf}(\tilde{w_i}^{(t)}, d)},
\end{equation}
where tf-idf measure is computed with augmented frequency,
\begin{equation*}
\begin{split}
	\mathrm{tfidf}(w, d) &= \mathrm{tf}(w, d) \times \mathrm{idf}(w) = \\
	&= \Big( \frac{1}{2} + \frac{f(w, d)}{\max\limits_{w' \in d} f(w', d)} \Big) \log \frac{|D|}{|\{d' \in D: w \in d'\}|},
\end{split}
\end{equation*}
where $f(w, d)$ is the number of occurrences of token $w$ in document $d$. Usage of tf-idf score is preferred to co-document frequency, since the latter can be skewed towards tokens that commonly occur together but do not define interpretable topics, as was shown in \cite{nikolenko2017topic}.

Next modification is \textit{word embedding coherence}, which we define as
\begin{equation} \label{m:3}
	\mathrm{Coh_{we}}(t) = \frac{1}{n(n - 1)} \sum\limits_{\tilde{w_i}^{(t)} \neq \tilde{w_j}^{(t)}} d(v(\tilde{w_i}^{(t)}), v(\tilde{w_j}^{(t)})),
\end{equation}
where $v: \mathcal{W} \rightarrow \mathbb{R}^{d}$ is a mapping from tokens to $d$-dimensional vectors and $d: \mathbb{R}^{d} \times \mathbb{R}^{d} \rightarrow \mathbb{R}$ is a distance function. 

Another class of quality measures is based on the idea  of \textit{pointwise mutual information} (PMI) between top tokens in a topic. Following \cite{lau2014machine}, we survey three variations of this idea.

\begin{itemize}
\item the basic \textit{pairwise} PMI measure \cite{newman2010automatic}, defined as
\begin{equation} \label{m:4}
	\mathrm{PMI}(t) = \frac{2}{n(n - 1)} \sum\limits_{i=1}^{n-1}\sum\limits_{j=i+1}^{n} \log \frac{p(\tilde{w_i}^{(t)}, \tilde{w_j}^{(t)})}{p(\tilde{w_i}^{(t)}) p(\tilde{w_j}^{(t)})};
\end{equation}
\item the \textit{normalized} PMI measure \cite{bouma2009normalized}, defined as
\begin{equation} \label{m:5}
	\mathrm{NPMI}(t) = \frac{2}{n(n - 1)} \sum\limits_{i=1}^{n-1}\sum\limits_{j=i+1}^{n} \frac{\log \frac{p(\tilde{w_i}^{(t)}, \tilde{w_j}^{(t)})}{p(\tilde{w_i}^{(t)}) p(\tilde{w_j}^{(t)})}}{-\log p(\tilde{w_i}^{(t)}, \tilde{w_j}^{(t)})};
\end{equation}
\item the \textit{pairwise log conditional probability} (LCP) measure \cite{Mimno2011}, defined as
\begin{equation} \label{m:6}
	\mathrm{LCP}(t) = \frac{2}{n(n - 1)} \sum\limits_{i=1}^{n-1}\sum\limits_{j=i+1}^{n} \log \frac{p(\tilde{w_i}^{(t)}, \tilde{w_j}^{(t)})}{p(\tilde{w_i}^{(t)})}.
\end{equation}
\end{itemize}

Having defined measures (\ref{m:1})--(\ref{m:6}), let us compare them in terms of correlation with human judgement of topical quality. We borrow this comparative results from the recent paper by Sergey Nikolenko \cite{Nikolenko2016}, where an assessment experiment was conducted. Three different models, namely pLSA, LDA, and ARTM, were assessed, where each topic was described by its 20 most probable words. The human experts were asked the following question: ``Do you understand why these words are collected together in this topic?'', with three possible answers: (a) absolutely not; (b) partially; (c) yes. The measures \ref{m:1} -- \ref{m:5} were calculated as described above; for word embedding coherence a pre-trained word2vec \cite{mikolov2013distributed} model of dimension 500, trained on a large Russian language corpus \cite{ai2015evaluating, panchenko2018russe}, was used as mapping $v$ along with inverse cosine similarity as a distance function: $d(x, y) = 1 - \langle x, y\rangle$. Area under ROC curve (ROC-AUC) \cite{hand2001simple,ling2003auc}, where scores are given by the described measures and labels are manually assigned by human experts, is used for evaluation. The results of the experiment are shown in the table \ref{m:res}.

\begin{table}[htb]
\begin{tabular}{ c | c c c | c c c }
 Model & \multicolumn{6}{c}{Quality measures} \\ 
\hline 
& $\mathrm{Coh}$ & $\mathrm{Coh_{tfidf}}$ & $\mathrm{Coh_{we}}$ & $\mathrm{PMI}$ & $\mathrm{NPMI}$ & $\mathrm{LCP}$ \\
\hline
pLSA & 0.7720 & 0.8910 & \textbf{0.8954} & 0.8675 & 0.8707 & 0.8811 \\  
LDA & 0.7817 & 0.8748 & \textbf{0.8786} & 0.8469 & 0.8372 & 0.8541   \\
ARTM & 0.7513 & 0.8439 & \textbf{0.8543} & 0.6637 & 0.6973 &  0.7738 
\end{tabular}
\centering
\caption{Comparison results between flat topic quality measures: area under curve (AUC) between human-assessed evaluation and automatic quality measures, borrowed from \cite{Nikolenko2016}. The best results for each model are highlighted in \textbf{bold}.} \label{m:res}
\end{table}

We see that the word embedding coherence is in larger agreement with human assessors than other measures. The reason to this might be that $\mathrm{Coh_{we}}(t)$ measures \textit{paradigmatic relatedness} \cite{schutze1993vector} of tokens, as underlying word2vec model $v$ is trained to make words with similar contexts have similar vectors, and it better approximates human understanding of word similarity than \textit{syntagmatic relatedness}, measured by cooccurrence- and tfidf-based measures.

\section{Visualization of topic models}\label{review:viz}
Since its inception, topic modeling has been successfully used for visualizing and navigating through large scientific corpora. Oftentimes, when introducing a topic model with new properties, authors provide a convenient visualization of their models. In their seminal work, Blei et al. \cite{blei2006dynamic} proposed a dynamic topic model to track evolution of topics over time. They demonstated it by building a dynamic topic model over a collection of Science journal articles from 1880 through 2000. Another interesting visualization system, RoseRiver \cite{cui2014hierarchical}, provides a flow visualization of hierarchical topic models of real-world news data. A more intuitive way to navigate topical hierarchies~---~a Sunburst chart ~---~is used in Hi\'erarchie \cite{smith2014hiearchie}, an hLDA-based framework.

\begin{figure}
\includegraphics[width=1.0\textwidth]{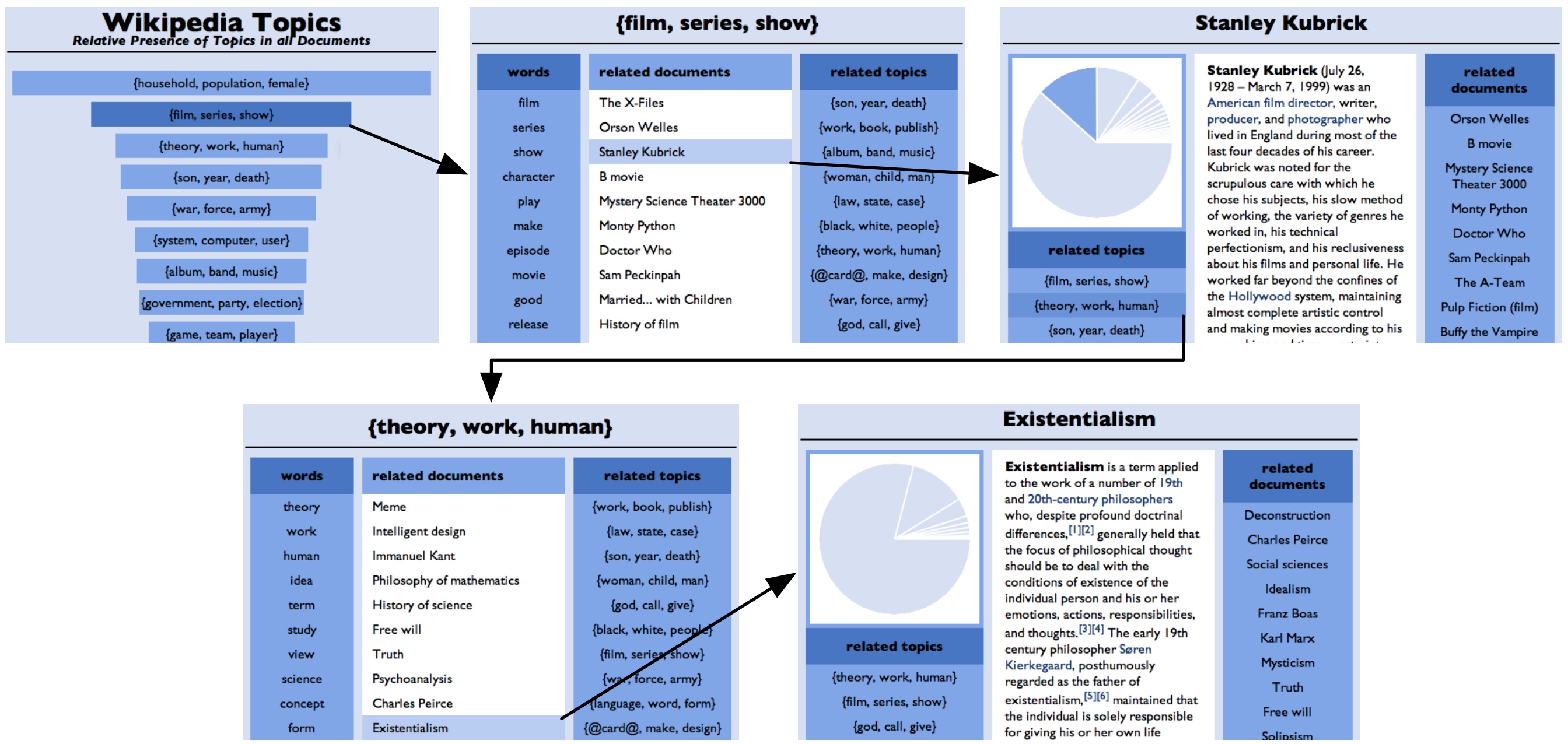}
    \caption{\label{fig:tmve}Topic Model Visualization Engine (TMVE). On the front page, a user can choose a generic topic, then display all documents relevant to it. From an article screen, the user can also choose from relevant documents and/or relevant topics, thus easily navigating around an area of interest. Their navigator is accessible online at http://bit.ly/wiki100.}
\end{figure}

Generic topical navigators for displaying flat topic models have been proposed as well. In 2012, Chaney et al. \cite{Chaney2012} proposed a Topic Model Visualization Engine (TMVE), a user-friendly system to navigate over English Wikipedia collection. A demonstration of its capabilities is given on Figure \ref{fig:tmve}. Some systems, such as Termite \cite{Chuang2012}, LDAVis \cite{sievert2014ldavis} or Serendip \cite{alexander2014serendip} focus more on internal structure of topic models rather then creating a navigator over a collection of documents. Termite is a convenient tool to visualize $\Phi$ matrix of a flat topic model. It highlights the most probable terms for each topic and introduces a new \textit{saliency} measure to select most relevant terms in a topic. Similar \textit{relevance} measure is introduced in  \cite{sievert2014ldavis}, along with LDAVis system. Their system can be used for both deep analysis of each topic using most relevant keyword as well as for displaying inter-topical relations by measuring topical distance and performing multidimensional scaling to a flat map. Apart from these levels of abstraction, Serendip provides an additional level of close-ups to inidividual passages and words. It makes use of $p(t|d)$ and $p(t|d,w)$ distributions and, similar to Termite, displays distribution of topics for each document and, just as RoseRiver, tracks evolution of topics similar to RoseRiver, but on a document scale.

\begin{figure}
\includegraphics[width=0.85\textwidth]{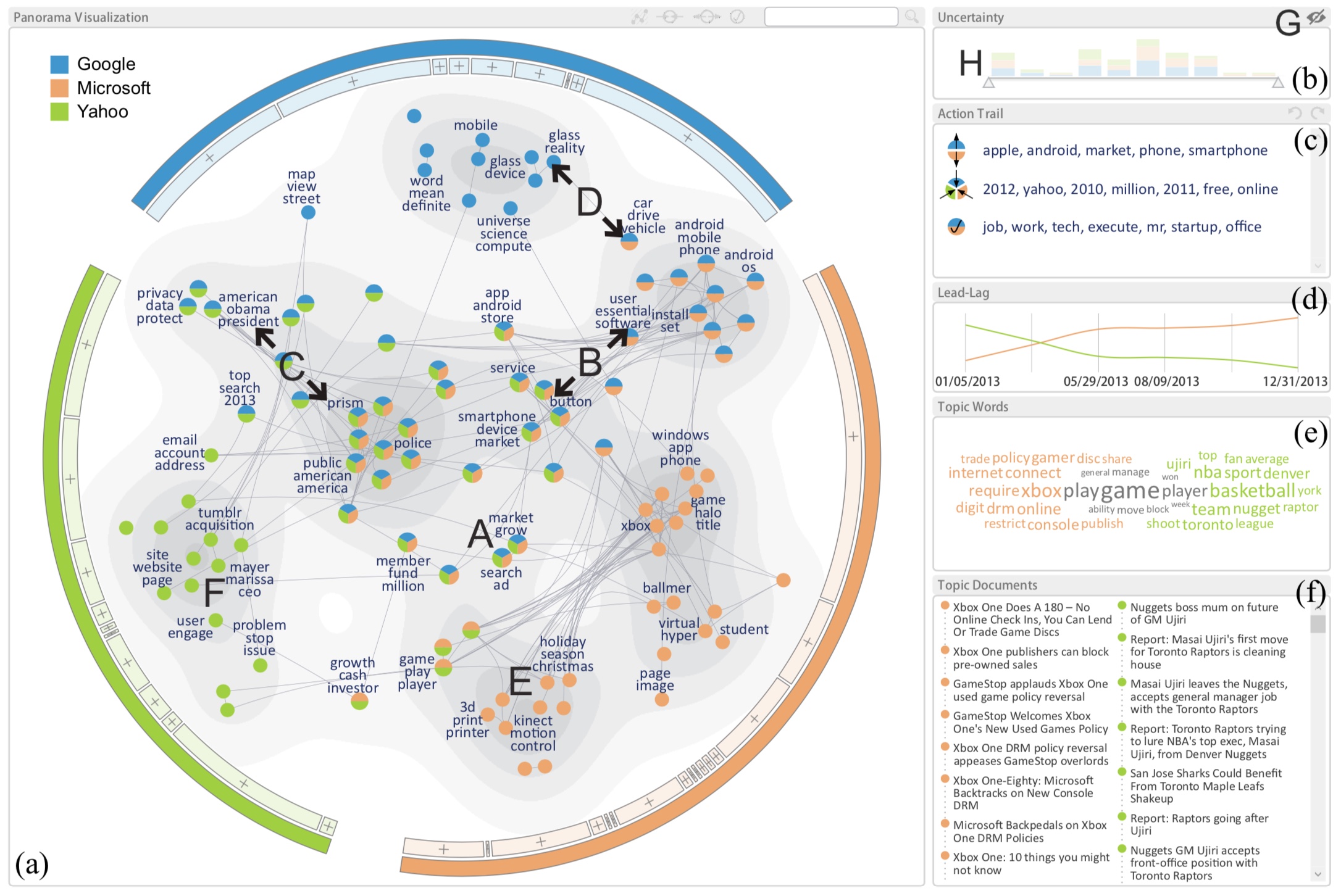}
    \caption{\label{fig:toppanorama}TopicPanorama. A dense and informative representation of topic model is available on the main screen. It includes: a) level-of-detail (LOD) visualization of topic inter-relations, d) lead-lag temporal dynamics between topics, (e-f) most relevant words and documents of each topic.}
\end{figure}

The idea to align topics and/or documents on a flat map and provide a meaning for distance on the map is ubiquitous in the topic modeling visualization field. TopicPanorama \cite{wang2016topicpanorama} uses circular graph representation and brings topics, discussed in the same document collections, together. The model is built on various data sources, including news articles, tweets, and blog data. Various levels of details are also displayed, which can bee seen on Figure \ref{fig:toppanorama}. Continuing the idea multi-collection visualization, Oelke et al. \cite{oelke2014comparative} present DiTop-View: a system of rich visualization of topical inter-relations. They propose \textit{topic coins} as a new way to display topics suitable for comparative analysis, and several heuristical criteria to determine topics that are specific for given collections. Gansner et al. \cite{gansner2012}, in a service they called TwitterScore, work with a single Twitter source. They propose a way to build a flat map with similar topics grouped into ``countries'', and a way to update this map in real time as the new data appears.

An interesting and relevant solution for topic alignment problem is proposed in \cite{fedoryakatechnology}. They formulate the problem as a search for a \textit{linear spectre} of topics, or a permutation $\pi$ over $|T|$ topics, such that sum of distances $\rho: T \times T \rightarrow [0, +\infty)$ between neighboring topics in this permutation $\rho(t_{\pi_{i}}, t_{\pi_{i+1}})$ is minimal:
\begin{equation}\label{fedoryaka}
	\sum\limits_{i=1}^{|T|-1}\rho(t_{\pi_{i}}, t_{\pi_{i+1}}) \rightarrow \min\limits_{\pi}.
\end{equation}
If $\rho(t_i, t_j)$ is expressed as a square matrix $R \in \mathbb{R}^{|T|\times|T|}$, where $R[i, j] = \rho(t_i, t_j)$, then the problem \ref{fedoryaka} can be formulated as a search for Hamiltonian path of minimal weight in a complete weighted undirected graph with weight matrix $R$, that can be solved either exactly (which is only feasible for small $|T|$) or approximately. A reasonable choice of intrinsic (that is, only depending on $\Phi$ matrix of a topic model) measures $\rho$, as well as usage of polynomial time Lin-Kernighan-Helsgaun algorithm \cite{helsgaun2000effective} to solve problem \ref{fedoryaka}, are proposed. Another notable contribution of \cite{fedoryakatechnology} is an open source VisARTM\footnote{\url{https://github.om/bigartm/visartm}} application, which supports automatic construction of topic models with required properties using BigARTM library. Different visualization modes are available, including temporal and hierarchical visualizations and in-depth textual statistics for each document in a collection.

Visualization of topic models is a rich field of study with an abundancy of recent contributions, of which we covered only a few. For a complete interactive review of modern visualization systems, see \cite{kucher2015text}. Another survey, in the Russian language, is presented in \cite{aysina2015review}.

\chapterconclusion
In Chapter 1, the research problem was stated and supported by the existing work. Probabilistic flat and hierarchical topic modeling, the area of our foremost contribution, was introduced. Major quality measures for flat topic models, which will form the basis of measures for hierarchical topic models, were described and compared. Multiple recent visualization frameworks for topic modeling were demonstrated, which supports the point of applicability of topic modeling for exploratory search engine construction.

\finishrelatedwork

\chapter{Proposed framework}

\section{Quality of topical edges}
In Section \ref{review:qual}, we surveyed classical measures of topical quality, which are in agreement with human understanding of what makes topic good or bad. We have also shown that these measures can be expressed as an average of functions $f(\tilde{w_i}^{(t)}, \tilde{w_j}^{(t)}), \tilde{w_i}^{(t)} \in W^{(t)}, \tilde{w_j}^{(t)} \in W^{(t)}$ depending on top tokens of the topic $t$.

However, a hierarchical topic model consists not only of its levels, but also of relations between topics from the neighboring levels, whereas the classical measures of the model's levels take these dependencies completely out of consideration. Hence, they fractionally depicts the quality of a \textbf{hierarchical model}. This section is aimed to bridge the gap by proposing several quality measures for the ``parent\,-\,child'' relations between topics in a hierarchical model. We propose two kinds of quality measures, \textit{extrinsic} and \textit{intrinsic}. The former will take advantage of word cooccurrences in external corpora, and the latter will only use internal parameters of a topic model, hence the names.

\subsection{Extrinsic similarity based measures}
We extend the classical evaluation scheme from Eq. \ref{review:qual_eq} to a hierarchical form
\begin{equation}
	\mathrm{Quality}_e(a, t) = \frac{1}{C} \sum\limits_{i=1}^{n}\sum\limits_{j=1}^{n} f(\tilde{w_i}^{(a)}, \tilde{w_j}^{(t)}),
\end{equation}
where $f(\tilde{w_i}^{(a)}, \tilde{w_j}^{(t)})$ is a measure of cooccurrence of top tokens $\tilde{w_i}^{(a)} \in W^{(a)}$ and $\tilde{w_j}^{(t)} \in W^{(t)}$, and $C$ is measure-specific norm. We refer to Section \ref{review:qual} and denote document frequency as $d(\tilde{w_i}^{(a)})$, co-document frequency as $d(\tilde{w_i}^{(a)}, \tilde{w_j}^{(t)})$, and vector mapping from token to real-valued $d$-dimensional vector as $v(\tilde{w_i}^{(a)})$. Then we define our measures as:
\begin{equation}\label{extr_measures}
\begin{split}
    \mathrm{EmbedSim}(a, t) &= \dfrac{1}{C}\sum\limits_{i=1}^n \sum\limits_{j=1}^n \left[ \tilde{w_i}^{(a)} \neq \tilde{w_j}^{(t)} \right] \cdot \langle v(\tilde{w_i}^{(a)}), v(\tilde{w_j}^{(t)})\rangle, \\
    \mathrm{CoocSim}(a, t) &= \dfrac{1}{C}\sum\limits_{i=1}^n \sum\limits_{j=1}^n \left[ \tilde{w_i}^{(a)} \neq \tilde{w_j}^{(t)} \right] \cdot \ln\dfrac{d(\tilde{w_i}^{(a)}, \tilde{w_j}^{(t)}) + \varepsilon}{d(\tilde{w_j}^{(t)})},
\end{split}
\end{equation}
where $C = \sum\limits_{i=1}^n \sum\limits_{j=1}^n \left[ \tilde{w_i}^{(a)} \neq \tilde{w_j}^{(t)} \right]$ is the number of distinct top tokens pairs.

To estimate (co-)document frequencies and to train mapping $v$, a large external corpus (e.g. Wikipedia or Twitter) is usually used.

\subsection{Intrinsic similarity based measures}
Another option for comparing parent and child topics is provided by a topic model itself: we can compare them as probability distributions. Two standard similarity measures for distributions P and Q are Hellinger distance and Kullback-Leibler divergence. The first one is a bounded measure and can be interpreted as distance between two topics in some space. The second is an unbounded asymmetric measure and can be interpreted as ``how much information will be lost if we substitute parent topic P with some child topic Q''.

\begin{equation}\label{intr_measures}
\begin{split}
    \mathrm{HellingerSim}(a, t) &= 1 - H(p(w|a), p(w|t)) = \\ &= 1 - \dfrac{1}{\sqrt{2}} \sqrt{\sum\limits_{i=1}^{|\mathcal{W}|} (\sqrt{\phi_{i a}} - \sqrt{\phi_{i t}})^2 }, \\
    \mathrm{KLSim}(a, t) &= -D_{KL}(p(w|a)\|p(w|t)) = \\ &= \sum\limits_{i=1}^{|\mathcal{W}|} \phi_{i a}\ln\dfrac{\phi_{i t}}{\phi_{i a}}.
\end{split}
\end{equation}
\section{Quality of topical hierarchy}
The goal now is to combine the edges measure into some construction, which would be a representative quality score for a hierarchy as a whole.

\paragraph{Normalization:} \label{note:normalization} Hereafter we work with a normalized matrix $\Psi$ as the following:

\begin{equation}
\label{eq:psi_normalization}
\tilde{\psi_{ta}} = \frac{\psi_{ta} - \min_t{\psi_{ta}}}{\max_t{\psi_{ta}} - \min_t{\psi_{ta}}}.
\end{equation}
It allows to apply shared topic-agnostic threshold and to rank all values of $\Psi$ matrix on the same scale.

\subsection{Averaging quality}
In the spirit of \cite{Mimno2011} where the average coherence was used as a model quality measure, let us consider the average edge quality as quality measure for hierarchical model ($\Phi, \tilde{\Psi}$):
\begin{equation}
	\mathrm{AvgQuality}_Q(k) = \dfrac{\sum\limits_{a, t \in A \times T} \left[ \tilde{\psi_{t a}} > k \right] \cdot Q(a, t)}{\sum\limits_{a, t \in A \times T} \left[ \tilde{\psi_{t a}} > k \right]}.
\end{equation}
The particular hierarchy configuration depends on the chosen threshold $k$ for $\tilde{\Psi}$, which determines what probability $p(t|a)$ is sufficient to include an edge connecting $t$ and $a$ into the hierarchy.  Therefore different thresholds lead to different values of a quality measure.

\subsection{Ranking quality}
Another approach to form a quality measure with an interpretable value is to consider the process of establishing a hierarchy as a ranking process. Consider that we have built a model i.e. we have matrices $\Phi$, $\Theta$, and $\tilde{\Psi}$ for each level. It would be natural to accept only the most meaningful edges according to a human's point of view. As our edge measures turned out to be good approximators of the assessors' judgment, we can choose only $k$ edges with the top scores of some fixed measures. If our model is ``good'', then top-$k$ scored edges (let us call them ``the request'') should match with top-$k$ maximal elements of the $\tilde{\Psi}$ (let us call them ``the response''). The difference between the request and the response for each $k$ was evaluated by common ranking measures, such as:

\begin{itemize}
\item Average Precision @ $k$ -- described in \cite{li2011short},
\item Normalized Discounted Cumulative Gain (NDCG) @ $k$ -- described in \cite{li2011short},
\item Inverse Defect Pairs (Inverse DP) @ $k$ -- the inverse value of the number of pairs that appear in the wrong order (i.e. are reversed) in the response.
\end{itemize}

\section{Heterogeneous topic modeling}\label{algos}

Let us denote $Train(D_{train}, \Phi_{init})$ an iterative algorithm which constructs a hierarchical hARTM model over collection $D_{train}$ with initial approximation $\Phi_{init}$. We propose two meta-algorithms, \texttt{concat} and \texttt{heterogeneous}, which re-use $Train$ procedure to build a hierarchical topic model for heterogeneous collections $\left\{ D_i \right\}, i = 1, \textellipsis, n$.

\subsection{\texttt{concat} algorithm}
In this algorithm, we simply concatenate all document collections and build a topic model upon it.
\begin{enumerate}
\item Concatenate all collections $D_{train} \leftarrow \bigcup\limits_{i=1}^{n} D_i$;
\item Build hARTM model $\Phi \leftarrow Train(D_{train}, 0_{w t})$;
\item Return $\Phi$.
\end{enumerate}

\subsection{\texttt{heterogeneous} algorithm}
In this algorithm, without loss of generality, we denote the first document collection $D_1$ as the \textit{base collection}. Collections $D_2, D_3, \textellipsis$ will be denoted as \textit{new collections}. Additionally, let $\Phi_0$ denote the initial model, built upon base collection and provided as a parameter. Our algorithm works in iterative fashion, on each iteration $i$ improving the current model $\Phi_i$.

\begin{enumerate}
\item Initialize train set $D_{train} \leftarrow D_1$;
\item 
\textbf{For each} $i = 1, \textellipsis, N$:
\begin{enumerate}
\item Form a new small batch $D_{b_i} \leftarrow Sample(\{D_2, D_3, \textellipsis\})$;
\item Add new batch to train set $D_{train} \leftarrow D_{train} \cup D_{b_i}$;
\item Build hARTM model $\Phi_i \leftarrow Train(D_{train}, \Phi_{i-1})$;
\end{enumerate}
\item Return $\Phi_N$.
\end{enumerate}

\section{Exploratory search engine}
To demonstrate the quality of \texttt{heterogeneous} algorithm, as well as to prove applicability of hierarchical topic modeling for creating exploratory search engines, we have implemented \texttt{Rysearch} system. It is a web application with rich visualization, searching and article navigating facilities.

In the following subsections we will discuss its three key components, namely client-server architecture, user interface, and user experience, greater detail.

\subsection{Client-server architecture}
\begin{figure}
\includegraphics[width=1.0\textwidth]{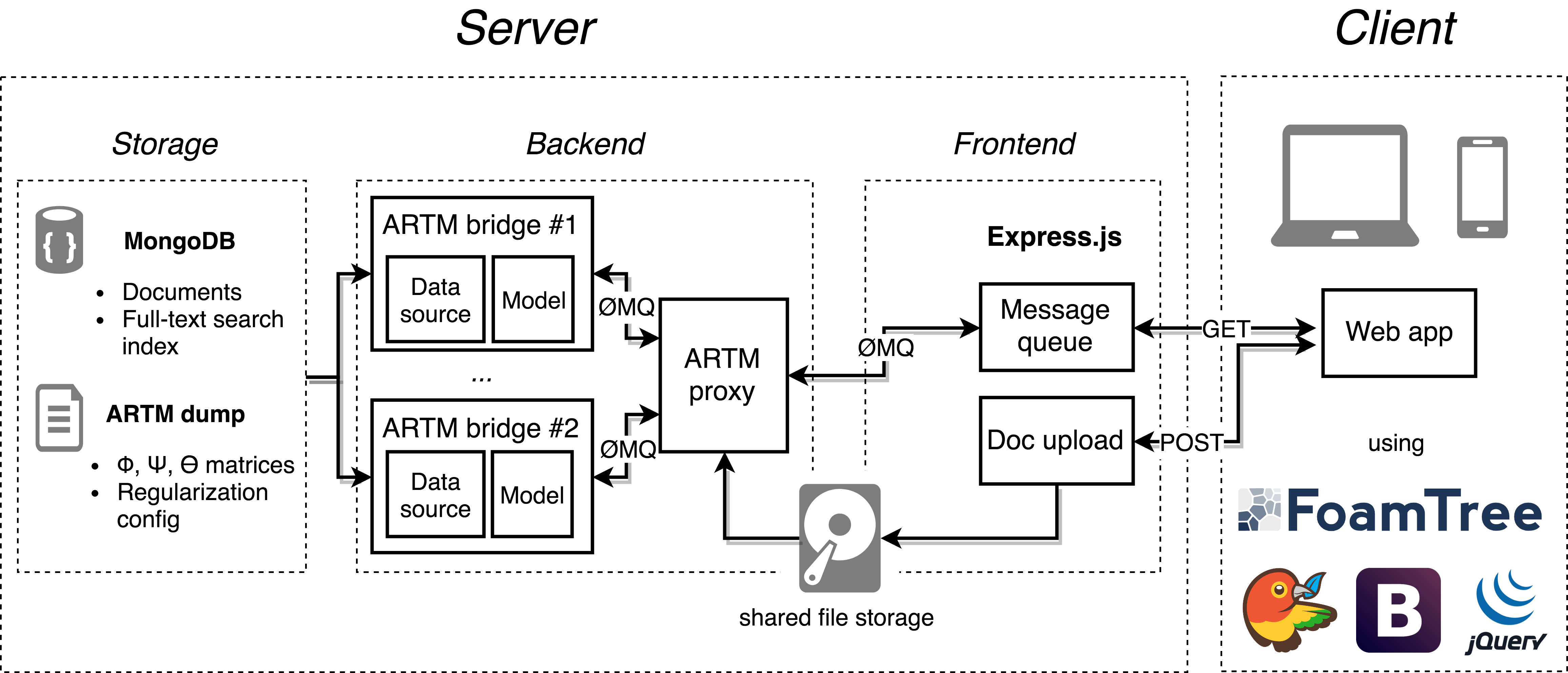}
    \caption{\label{fig:arch}Client-server architecture of \texttt{Rysearch}.}
\end{figure}
\texttt{Rysearch} has been designed as an experimental system, with the potential to scale for real life scenarios. Scaling, in this case, means being able to put up with high work load from two ends: larger streams of data coming from multiple heterogeneous sources, and increasing number of simultaneous end-users. In such setup, application design should be highly modular, with small independent blocks instead of a single monolith architecture. Our \textit{server side} meets these demands in the following ways.

Firstly, its \textit{storage} facilities make heavy use of MongoDB database and its features, such as advanced full-text indexing and extensible structure of MongoDB documents.

Secondly, its \textit{backend} architecture is designed hierarchically, with \textit{ARTM proxy} module receiving frontend queries and balancing them evenly between workers, or \textit{ARTM bridges} (these modules are called bridges as they were designed as thin clients that provide a ``bridge'' between BigARTM library and frontend servers). If one ARTM bridge becomes idle, it receives a new portion of work in a round-robin fashion. If an ARTM bridge stops responding to periodic \texttt{ping} queries, it is considered faulty and is excluded from round robin. Inside ARTM bridge are two database connectors, \textit{DataSource} and \textit{Model}, which handle requests to MongoDB and ARTM model databases, respectively. The former is used for displaying documents and full text searching. The latter is used for displaying hierarchical topic map, retrieving documents of a particular topic and for document searching. 

Thirdly, all backend and frontend modules of \texttt{Rysearch} are implemented as \textit{microservices}, that is, yhey are loosely connected via ZeroMQ \cite{hintjens2013zeromq}, or {\O}MQ, a protocol for asynchronous messaging without a dedicated message broker. This architectural style is associated with increased deployability, modifiability and resilience to design erosion, according to \cite{chenmicroservices}.

Being a microservice-based application, \texttt{Rysearch} also benefits from a possibility to write services in different programming languages, as well as to choose the most suitable frameworks for each service. As such, all backend services are written in Python, using pyzmq library and python wrapper for BigARTM. Frontend service is written is Javascript, basing on node.js platform with node package manager (npm) and express.js framework for serving client requests (using GET and POST methods of HTTP protocol).

The main befenit of node.js / express.js framework is that is was designed to serve requests in asynchronous fashion using a single-threaded non-blocking event loop. This is very useful for building fast application with high throughput. When the main screen of \texttt{Rysearch} is being loaded, a single query (``show me the topic map'') is sent to the frontend, which serves the reply from its internal cache, without sending queries to the backend. All the heavy computation that cannot be pre-computed is forwarded to backend workers using a proxy, as we described earlier, but when possible, responses are served using fast and high-throughput frontend servers. Another benefit of these frameworks is an ability to communicate asynchronously with backend, using zmq binding for node.js, without any hassle. If we served user queries in a synchronous manner, say using a thread pool of workers, we would have to do extra job to synchronize communication between our frontend and a backend proxy, which could be error-prone and unstable.

A client side is written in Javascript, uses Bootstrap and jQuery libraries, and is maintained by Bower package manager. To communicate with the server side, it sends asynchronous JSON-formatted (AJAX) requests and asynchronously waits for the responses in the same format. All messages, except an incoming document upload, are sent with GET method. Typically, a frontend server either sends the response back (using a previously cached response from the backend), or forwards client's request (with none to minimal preprocessing). In this case, a client ID is put into the \textit{message queue} and a message is sent to the backend proxy. When a reply from proxy arrives (or when the timeout of 1 minute is reached), a corresponding client ID is taken from message queue and the backend's response (again, with none to minimal postprocessing) is sent to the client. Uploading of a document (for document search, which we describe later) from a client is done using XMLHttpRequest level 2 \cite{xhr2}, or XHR2, which provides an API to transfer files from a client to a frontend server. It is then stored in a temporary disk storage, shared between frontend and backend servers, and a link to is sent via {\O}MQ channel. It is inconvenient to set up a shareable storage between servers just for uploading files, and in future versions we aim to come up with different scheme for processing file uploads.

\subsection{User interface}
\texttt{Rysearch} is single-page web application, meaning that it loads a single HTML page and dynamically updates the page as user interacts with the app. In each state of user interaction, the application can display one of two screens, or \textit{views}: map view or document view.

\subsubsection*{Map view}
\begin{figure}
\includegraphics[width=1.0\textwidth]{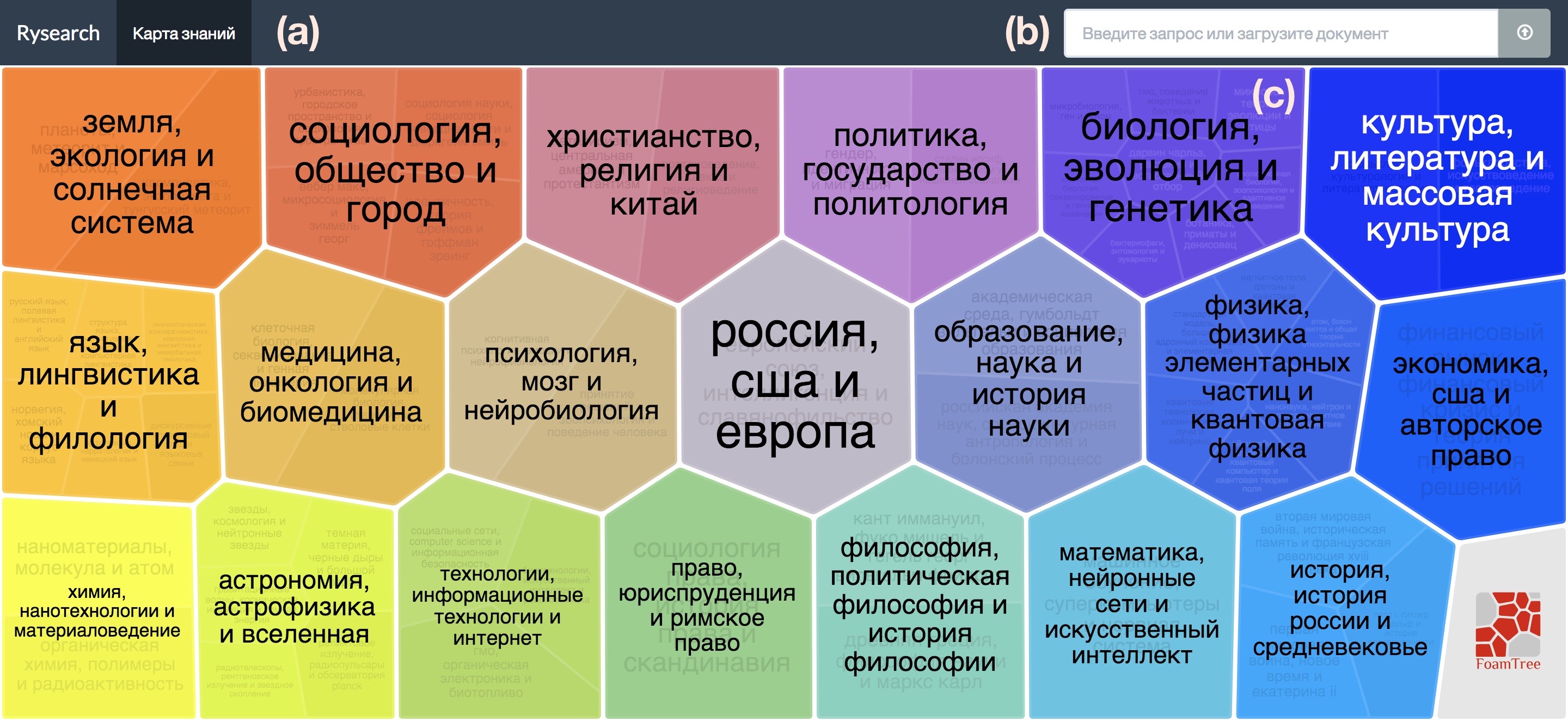}
    \caption{\label{fig:main}The map view of \texttt{Rysearch} user interface. On top of the page there are \textbf{(a)} navigation bar and \textbf{(b)} search widget, where users type in text queries or upload documents. Most of the page is occupied by \textbf{(c)} hierarchical map made out of tiles, which represent topics and their respective subtopics.}
\end{figure}

\begin{figure}
\includegraphics[width=1.0\textwidth]{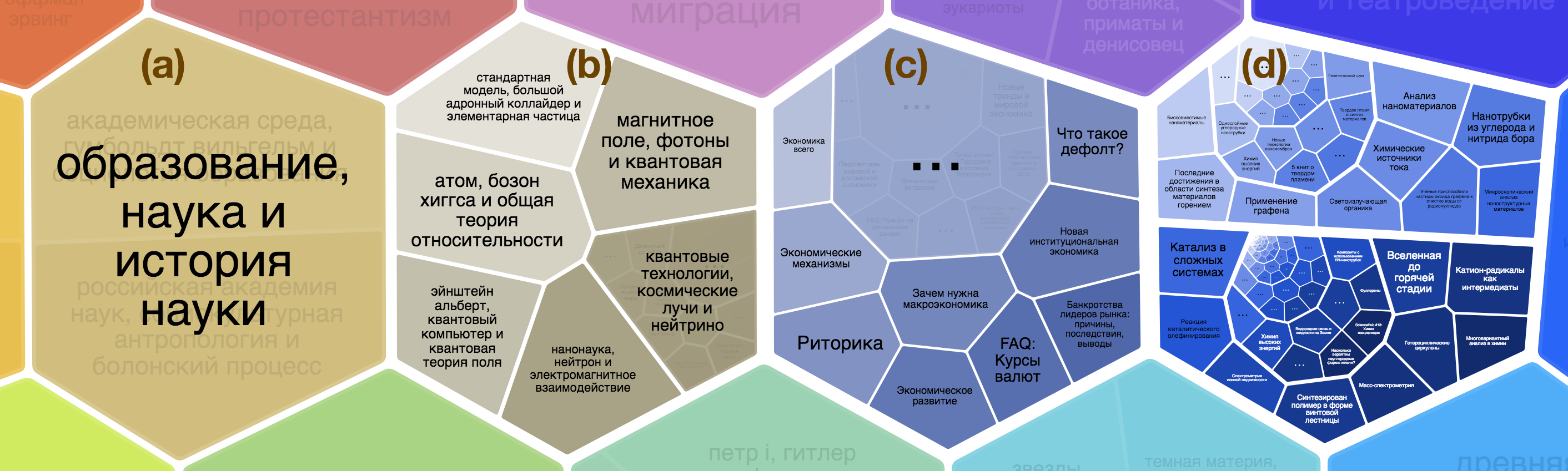}
    \caption{\label{fig:lod}Different levels of detail of topical cells on the map. By default, each topic is seen in the state \textbf{(a)}, which is a superficial view of top-3 tags associated with each top-level topic. By clicking on the cell this view is changed into \textbf{(b)}, where more detailed view of top-3 tags of each sub-topic is displayed. Another click changes cell into state \textbf{(c)}, where we see top-10 documents of a selected sub-topic. By clicking on (\textellipsis) tile, this cell can be infinitely deepened, resulting in an ``infinite scroll'', which is seen at the bottom sub-topic of a cell \textbf{(d)}.}
\end{figure}

On this screen, a user is presented with an overview of a whole document collection of an exploratory search engine. Documents are organized into topics, and topics are organized into sub-topics, in a one-to-many fashion: each document or sub-topic can be part of several sub-topics and topics, respectively. To find information of interest, a user can either zoom in to a desired \textit{level of detail} (see Figure \ref{fig:lod}), or use a \textit{search widget}. We will explore searching facilities later in the next section, but for now we not that the search can go either by typing text queries, as in classical search engines, or by uploading a document, which is specific to topic model-based search.

If, after performing a search or after exploring some topic in great depth, a user wishes to go back, he can do so by following \textit{breadcrumbs} on a navigation bar. As a user opens a new tile, searches something, or opens a document, a new link is created there to the previous state on the map to allow to go back in the exploration process.

To display an interactive topical map, we use client-side FoamTree visualization library\footnote{\url{https://carrotsearch.com/foamtree/}}. It takes tree-like structures (such as topic-subtopics-documents structure, where one-to-many relations are duplicated across the levels of hierarchy) as an input and produces an interactive tiled map by iteratively constructing hierarchical Voronoi diagrams, embedded either into the whole screen on into cells of higher level. As FoamTree allows to specify approximate positions of high-level cells on the map by submitting initial ordering of Voronoi tiles, we make use of linear spectres, discussed in Section \ref{review:viz}, to bring similar high-level topics as close as possible on the map.

\subsubsection*{Document view}
\begin{figure}
\includegraphics[width=1.0\textwidth]{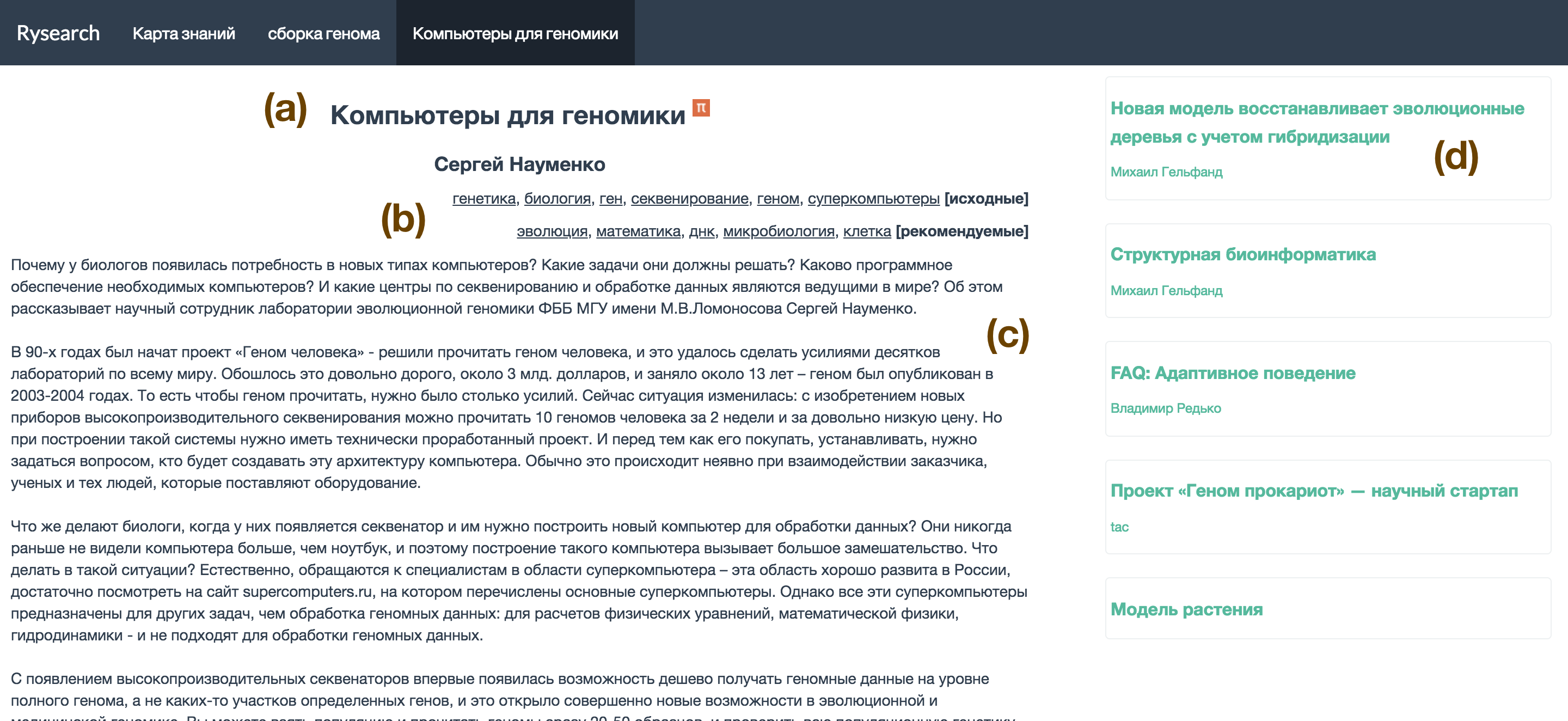}
    \caption{\label{fig:doc}The document view of \texttt{Rysearch} user interface. There are  \textbf{(a)} document title and author, as well as an indicator from each source this document comes from. For collections where documents are tagged by authors we display \textbf{(b)} the original tags, as well as top-5 recommended tags. Below is, of course, \textbf{(c)} document text, and to the right are \textbf{(d)} top-5 recommended similar documents.}
\end{figure}
On this view, a document meta-information, contents, and suggestions are presented. Meta-information includes the title, the author's name, and the icon representing the specific collection the document belongs to. Contents is a textual representation of a document, without any images or markup. Suggestions include suggested tags and suggested documents. The former can be helpful for users, who wish to perform a full-text search on \texttt{Rysearch} or elsewhere, but don't know which specific keywords to use. The latter provides another, horizontal level of navigation: suppose a user wants to explore documents similar to the given, but does not want to go up and down cells on the map view. Then he or she can use suggested documents as an alternative way of exploring the area of interest.

\subsection{User experience}

\begin{figure}
\includegraphics[width=0.75\textwidth]{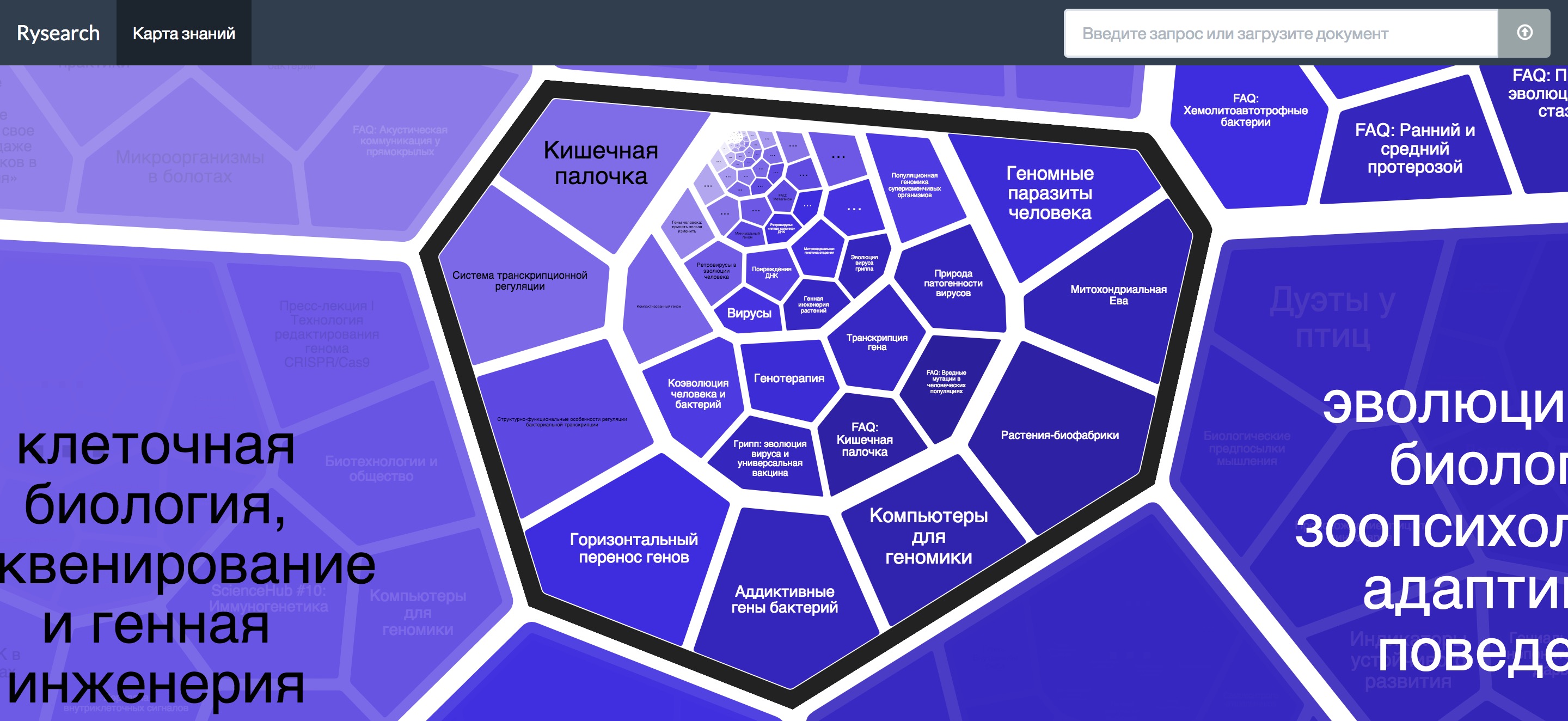}
    \caption{\label{fig:infs}Scenario 1: a user explores a known area of interest. A set of sub-topics and documents, most relevant to that topic, appear as they click on that topic. If they are dissatisfied with the presented results, they go deeper into (sub-)topic, which is possible because of the ``infinite scroll'', or wider to other topics, as similar topics are aligned closer on the map thanks to linear spectres. Therefore, to explore an area interest, either in-depth or superficially, a user does not typically have to jump over distant parts of map or over different browser windows -- a typical search query is concentrated in small region.}
\end{figure}

\begin{figure}
\includegraphics[width=0.75\textwidth]{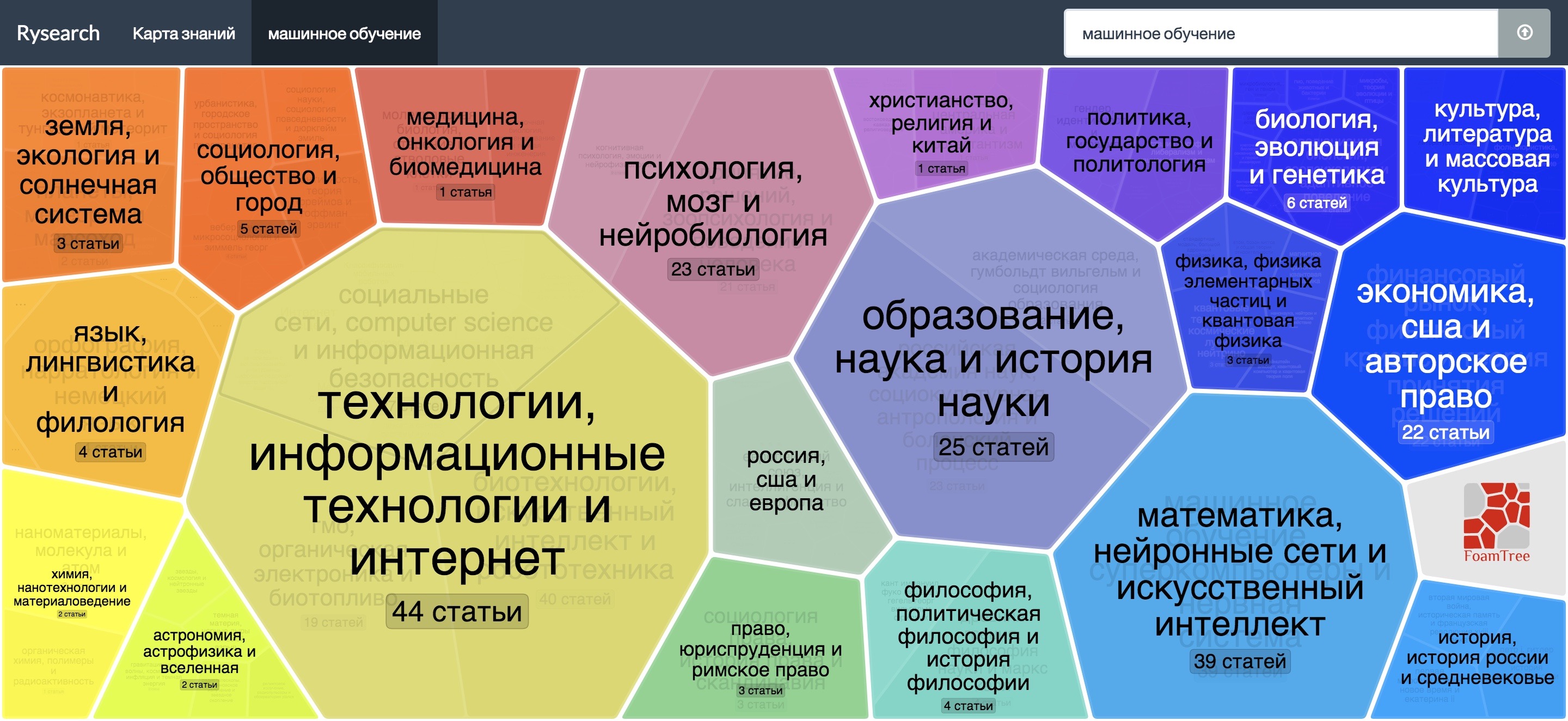}
    \caption{\label{fig:textsearch}Scenario 2: a user wishes to discover areas of interest by typing in text query. Immediately as they type a search query, a map is being dynamically rebuilt, thanks to asynchronous architecture and dynamic nature of FoamTree visualization library. A set of most important topics (up to 5 of them) is highlighted with tile size, with an additional identifier of a number of docs inside each topic, which are relevant to the exact text query. A user's goal is instantly achieved, even before they finish typing their query. They can also read specific documents that matched the query, as these documents will be highlighted in bold in ``infinite scroll'' view on the map.}
\end{figure}

\begin{figure}
\includegraphics[width=0.75\textwidth]{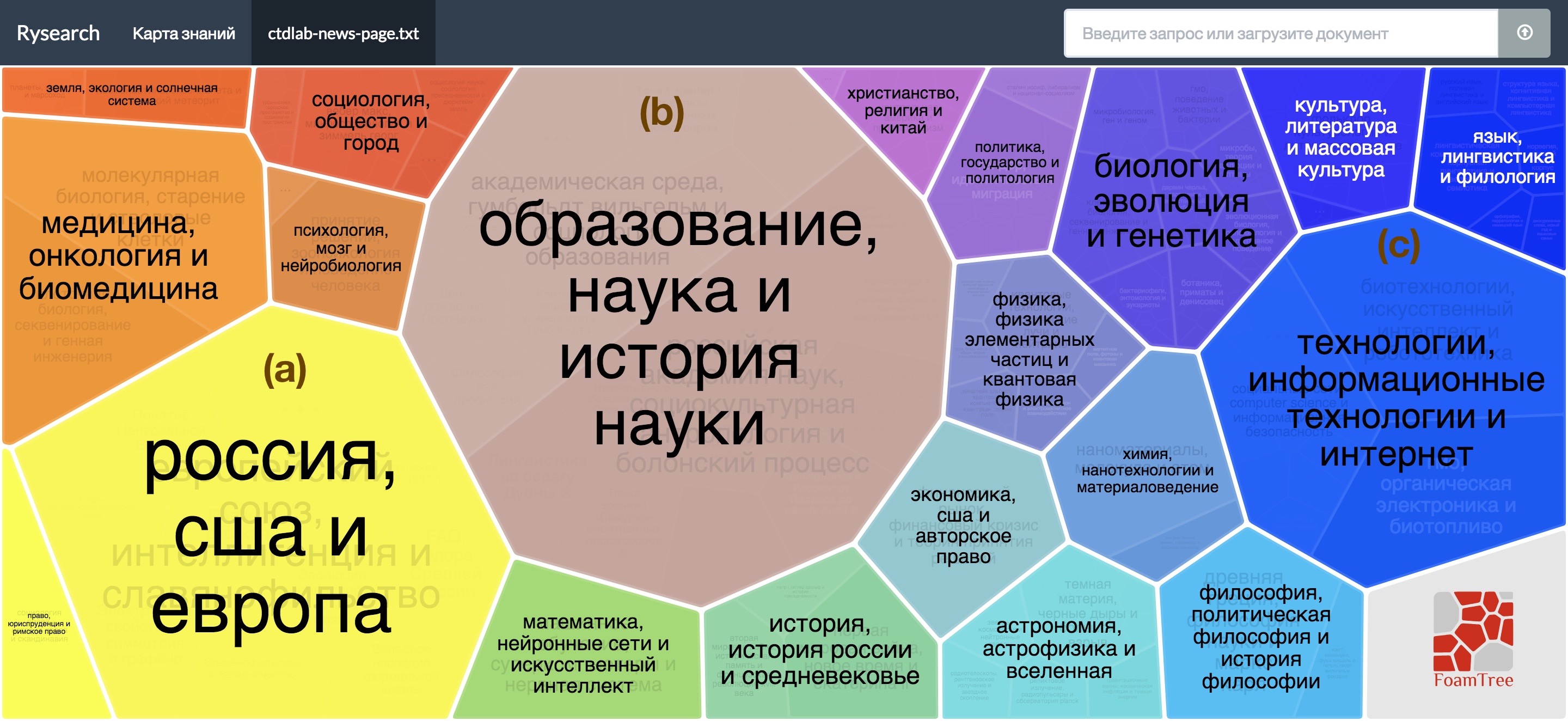}
    \caption{\label{fig:upload_short}Scenario 3a: a user wishes to discover areas of interest by uploading a document, which superficially discusses multiple topics. Suppose a user wants to discover what is an International laboratory ``Computer technologies'' by uploading their news web page to \texttt{Rysearch}. The resulting highlighted topics discuss \textbf{(a)} international affairs of countries, \textbf{(b)} science and education, and \textbf{(c)} information technology. The user, in this case, will have a correct understanding of the laboratory in a matter of seconds.}
\end{figure}

\begin{figure}
\includegraphics[width=0.75\textwidth]{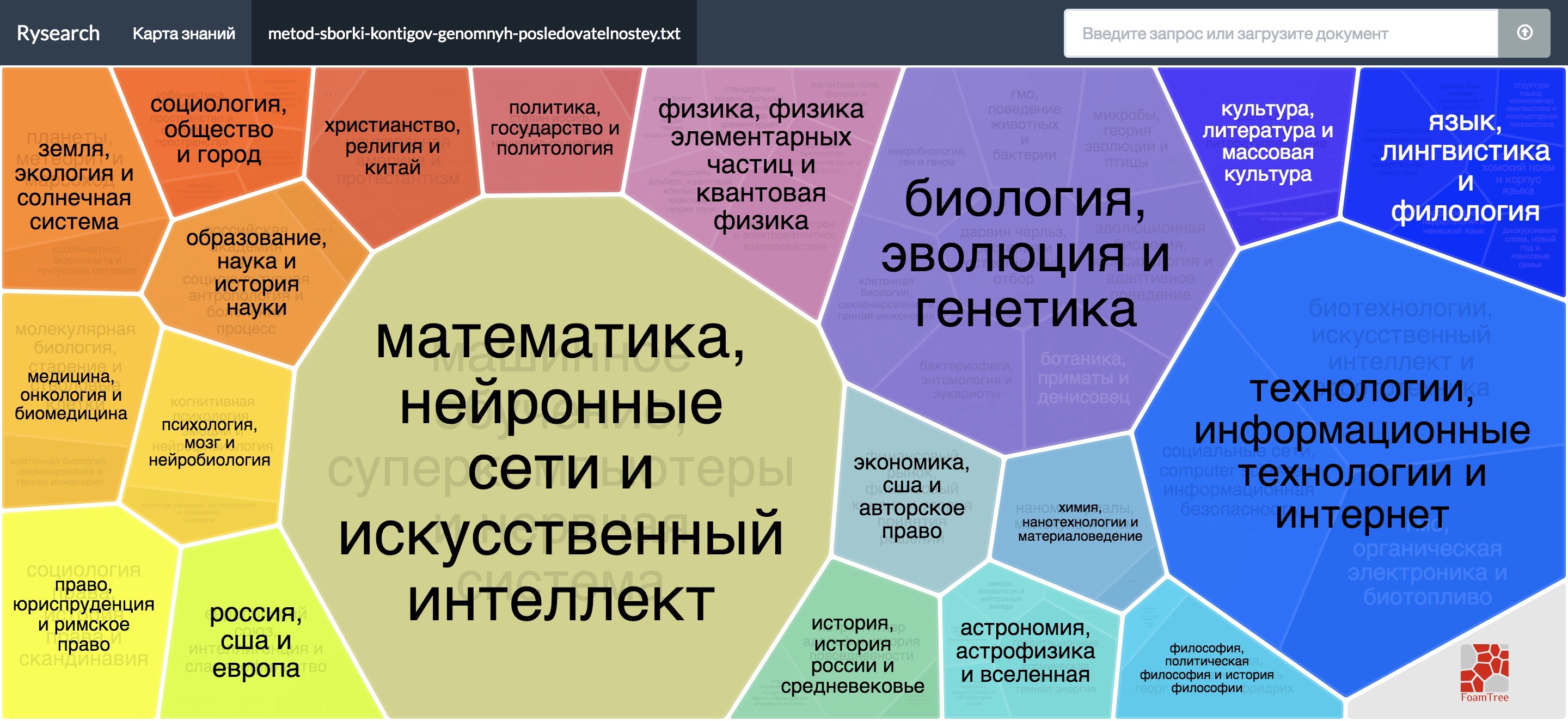}
    \caption{\label{fig:upload_long}Scenario 3b: a user wants to discover areas of interest by uploading a scientific article, which makes a focused contribution to a few scientific fields. He uploads a paper \cite{aleksandrov2012method}, which contributes to the field of bioinformatics by using methods from mathematics and computer science. Again, in this case, the engine was able to highlight all major areas of interest, without requiring the user to even open the article.}
\end{figure}

In this subsection, we explore a set of typical goals users can have while using the suggested search engine, and demonstrate that they indeed achieve their goals by referring to the user screens. We present these demonstrations on Figures \ref{fig:infs}, \ref{fig:textsearch}, \ref{fig:upload_short}, and \ref{fig:upload_long}.

\chapterconclusion
In Chapter 2, we introduced quality measures for topical edges of hierarchical models, arguing that previously proposed measures only fractionally depict the quality of HTMs. For this measures need to be aggregated to achieve a single quality score of a hierarchical model, we proposed two approaches of such aggregation: averaging and ranking. The former is simpler and more intuitive, while the latter is more interpretable. We also proposed two algorithms for building a topic model over heterogeneous sources, \texttt{concat} and \texttt{heterogeneous}. Finally, we described in detail an exploratory search engine, which make use of the proposed algorithm to build an interpretable hierarchical map of popular scientific topics.

In the following section we will show the agreement between the proposed measures and human judgement on topical edge quality and compare the proposed algorithms with the these measures.

\chapter{Experiments}
In this chapter, we describe experiments and data that were used to validate the proposed framework from Chapter 2. To validate the measures, we need to relate them to human experts' judgement of the topical edges quality, as we ultimately need to predict how much an arbitrary HTM is comprehensible and convenient for navigation by its end users. To validate \texttt{heterogeneous} algorithm, we will compare it to less elaborate \texttt{concat} algorithm using the proposed measures. In our experiments we use BigARTM \cite{vorontsov2015bigartm}, an open source library for topic modeling of large collections, and a web-based environment Jupyter Lab based on IPython for interactive experimentation and computing.

\section{Datasets}
To construct \mbox{``parent-child''} topic pairs for human annotation, we trained three two-level hierarchical topic models on three datasets:
\begin{itemize}
\item \texttt{Postnauka}, a popular scientific website with edited articles on a wide spectrum of topics, focusing on human sciences,
\item \texttt{Habrahabr} and \texttt{Geektimes}, social blogging platforms specializing in Computer science, engineering and IT entrepreneurship,
\item \texttt{Elementy}, a popular scientific website with a particular focus on life sciences.
\end{itemize}

\begin{table}
\begin{tabular}{l|r|r|r|r|r}
       & $|D|$ & $|W^{word}|$ & $|W^{tag}|$ & $|T_1|$ & $|T_2|$ \\
\hline
\texttt{Postnauka} & 2976  & 43196  & 1799 & 20 & 58 \\
\texttt{Habrahabr} & 81076 & 588400 & 77102 & 6 & 15 \\
\texttt{Elementy} & 2017  & 40452  & -- & 9 & 25 \\
\end{tabular}
\caption{\label{table:tm_datasets}Dataset parameters. $|D|$ is a collection size, $|W^{word}|$ is a number of unique words in the collection, $|W^{tag}|$ is a number of unique tags, $|T_1|$ is a number of first level (parent) topics, $|T_2|$ is a number of second level (child) topics.}
\end{table}

The first two collections, \texttt{Postnauka} and \texttt{Habrahabr}, were also used in constructing search index for \texttt{Rysearch}. Detailed characteristics of these datasets and corresponding topic models are provided in Table \ref{table:tm_datasets}.

To provide a mapping $v(w_i)$ from words to vectors for EmbedSim measure (Eq. \ref{extr_measures}), we used RusVectores \cite{kutuzov2016webvectors} pre-trained model, which was trained on external Russian National Corpus and Russian Wikipedia (with 600 million tokens, resulting in more than 392 thousand unique word embeddings). To measure document frequency $d(w_i)$ as well as co-document frequency $d(w_i, d_j)$ for CoocSim measure (Eq. \ref{extr_measures}), we used \texttt{Habrahabr} and \texttt{Postnauka} collections, as co-document frequency is difficult to compute on larger external corpora.

\section{Assessment experiment}
\subsection{Assessment task statement}
In the assessment experiment, conducted on Yandex.Toloka crowdsourcing platform, the participating experts were asked the following question: ``given two pairs of topics, $T_1$ and $T_2$,  decide whether one is a subtopic of another''. Possible answers were: ``$T_1$ is a subtopic of $T_2$'', ``$T_2$ is a subtopic of $T_1$'' and ``these topics are not related''. Topic $t$ was denoted by 10 top tokens from its probability distribution $p(w|t)$.

After the experiment was finished, the first two answers were grouped to denote a single answer ``these topics are somehow related'' as it was often difficult for assessors to distinguish between a parent and a child given their top tokens.

\subsection{Quality control}
The workers were selected from the pool of top-50\% Yandex.Toloka assessors, which takes into account the rating received during all previous annotation jobs completed by the assessor. Selected workers were required to undergo a training before entering an experiment. The training consisted of 22 pairs of topics, which we labelled manually.

Experts could have skipped some tasks if they were not sure. To ensure qualified responses during the experiment, we banned those workers who skipped more than 10 tasks in a row from participating in our experiment. To ensure the diversity of responses, we additionally allowed each worker to annotate not more than 125 edges each.

Upon successful completion of a task (which consisted of annotating 5 edges) each worker received \$0.02, or \$2.4 per hour on average.

\subsection{Experiment results}
Overall, 68 trusted workers participated in our study, each contributed around 100 assessed topical pairs. Assessment of one pair of topics, given their 10 top tokens, took around 5 seconds for each participants on average. Each topic pair was evaluated by at least five different experts, which gave us 6750 expert annotations for 1350 unique pairs.

Our participants were mainly from Russia and Ukraine, with age varying from 21 to 64 years.

\begin{figure}[t!]
\includegraphics[width = 1.0 \textwidth]{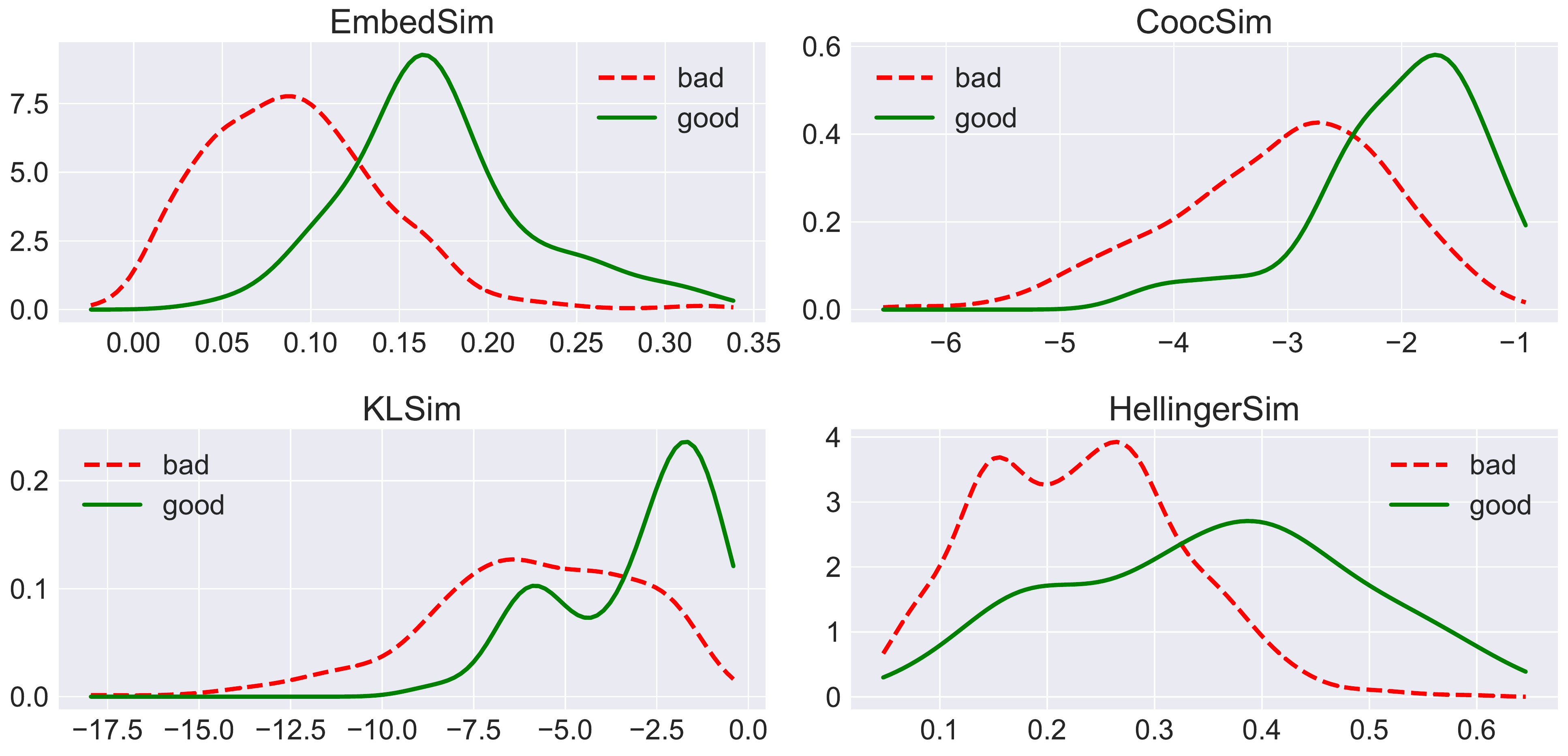}
\caption{\label{fig:mimno_like_graphs}Probability distributions of the proposed measure scores for ``bad'' and ``good'' edges from the assessment task.}
\end{figure}
\subsubsection*{Agreement}
\begin{table}[h]
\centering

\begin{tabular}{r|r|r}
Agreed assessors & Edge count & Edge percentage \\
\hline
3 & 374 & 27.7\% \\
4 & 468 & 34.7\% \\
5 & 508 & 37.6\% \\
\end{tabular}
\caption{\label{table:agreement}Inter-assessor agreement. For each pair of topics, we calculate how many assessors made the same verdict (that the topics from the pair are related or that they are not). For 5 assessors per pair, there is always a majority decision, but it can be reached by either 3, 4, or 5 assessors. In the second and the third column we show the quantity and the percentage of the edges with the number of agreed assessors from the first column.}
\end{table}

If many people think that there is a \mbox{``topic – subtopic''} relation between two particular topics in a model, a good measure should give a high score for such a pair of topics. In this case we say that a measure approximates assessors' opinion. Moreover, we want that measure to keep an order on the model edges consistent with this statement: the more people agreed that the relation exists – the higher the measure score should be.
In order to prove that the proposed measure holds this constraint, consider the following classification problem. Let us call \textit{the assessors' judgment} the fact that at least 4 of 5 assessors agreed that an edge exists in a hierarchy. If it holds, then assessors' judgment on this edge is equal to 1 (the edge is ``good''), and -1 (the edge is ``bad'') otherwise. In Table \ref{table:agreement} we see that for 72.3\% of assessed topical edges we can determine whether they are ``bad'' or ``good''. Let the edges of a hierarchical model be the objects: the positive and negative classes consist of the edges with a positive and a negative assessors' opinion respectively. Let the classifier based on the measure be the following:

\begin{equation}
\label{eq:classifier}
\mathrm{f}(t_1, t_2) = \mathrm{sign}(\rho(t_1, t_2) - w)
\end{equation}
where $t_1$ and $t_2$ are the topics from parent and child levels of the model respectively, $\rho$ is one of the proposed measures and w is a margin of the classifier. Having it written in this form, we can calculate ROC AUC for each classifier and estimate the quality of each measure: better approximators are expected to have better scores.

\begin{table}[h]
\centering

\begin{tabular}{l|r}
Measure      & Score \\
\hline
EmbedSim     &  0.878  \\
CoocSim      &  0.815  \\
KLSim        &  0.790  \\
HellingerSim &  0.766  \\
\end{tabular}
\caption{\label{table:measures_results}ROC AUC scores for the proposed measures.}
\end{table}

The Table \ref{table:measures_results} presents ROC AUC score for each classifier. One can see that the best classification quality was demonstrated by the classifier based on the EmbedSim measure (AUC = 0.878). The other measures demonstrated moderate yet acceptable consistency with the assessors' opinion: AUC values lied evenly above 0.75. 

For better understanding of this result one can see Figure \ref{fig:mimno_like_graphs}. For each graph the red line is a density distribution of the measure value for bad edges, and the green one is the same for good edges. The better some vertical line divides bad edges from good ones – the better the measure is. In further experiments we use the EmbedSim measure, as it demonstrated the best consistency with the assessors' judgment.

\subsubsection*{Interpretation}
\begin{figure}
\includegraphics[width=1.0\textwidth]{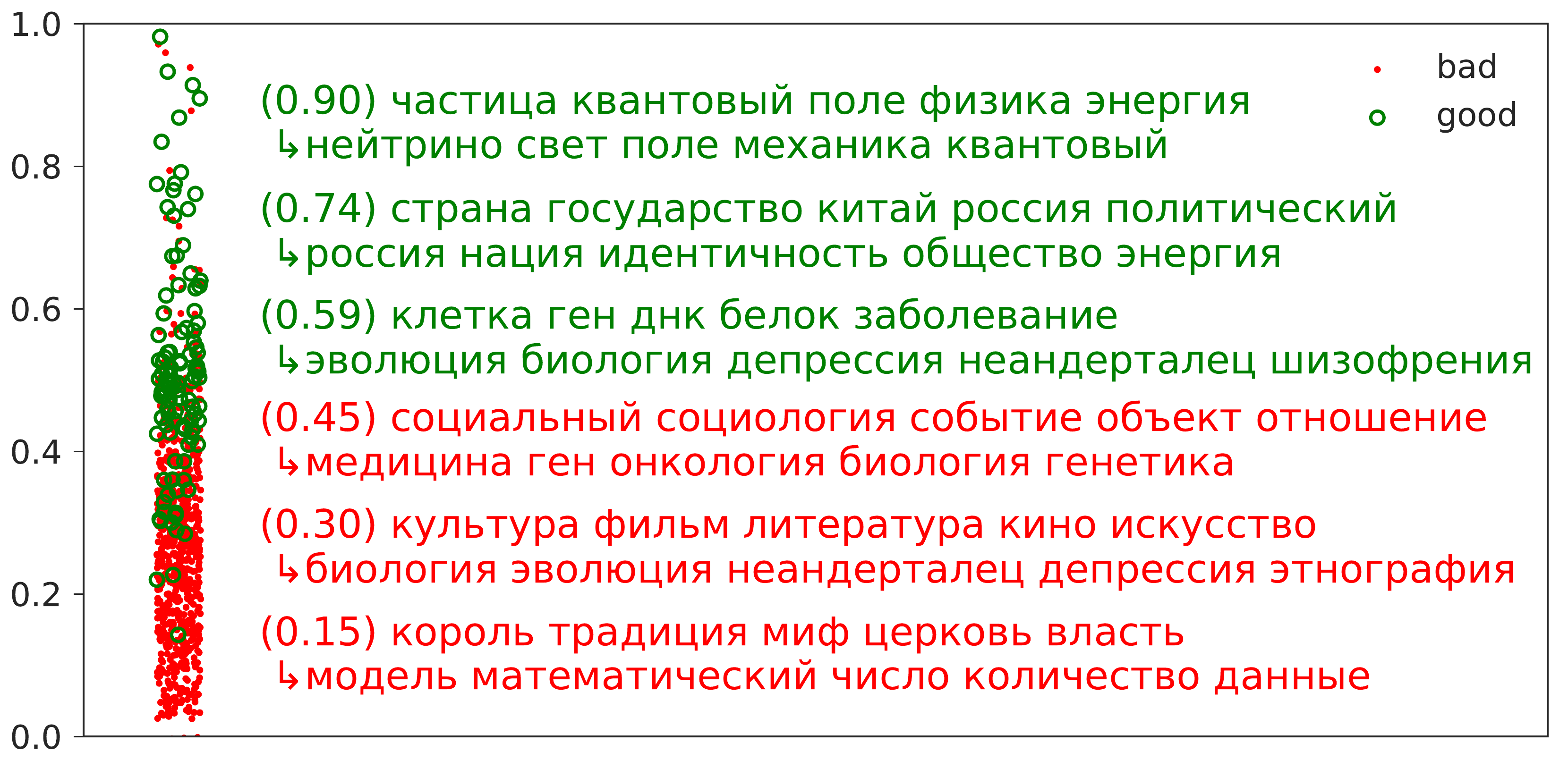}
\caption{\label{fig:embed_interp}Topics and their subtopics from the assessment task scored with the EmbedSim measure as hierarchy edges. Each topic or subtopic is represented by its 10 top tokens.}
\end{figure}

To understand how these measures work, let us consider an example. We are given six \mbox{``parent-child''} pairs of topics that were assessed by humans. Three of them are labeled as ``good'' (there is a semantic similarity between parent and child), other three are labeled as ``bad'' (little or no similarity). On Figure \ref{fig:embed_interp} one can see these pairs on the right, along with their scores given by the EmbedSim measure. The higher the score, the more confident the measure is. On the left there is a distribution of all the edges from an assessment task described in the following section. Y-coordinates of points are assigned according to measure score, and colors are set by the assessment experts.

\section{Comparison of \texttt{concat} and \texttt{heterogeneous} algorithms}

\subsection{Averaging measures}
\begin{figure}
\includegraphics[width=1.0\textwidth]{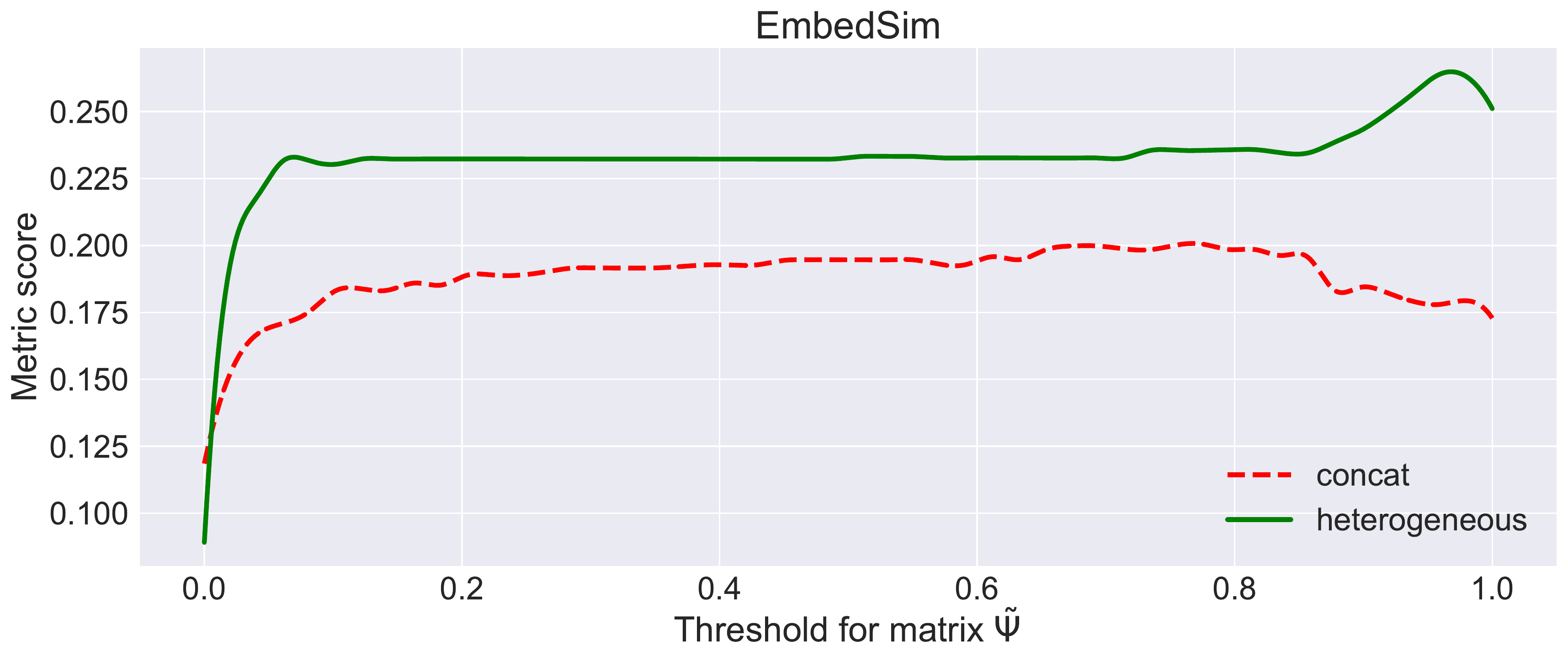}
\caption{\label{fig:averaging_measures}Averaging quality for the EmbedSim edge measure. The considered models (\texttt{concat} and \texttt{heterogeneous}) are described in Section \ref{algos}.}
\end{figure}

Figure \ref{fig:averaging_measures} depicts values of $\mathrm{AvgQuality}_{EmbedSim}(t)$ for all possible values of $t$. The \texttt{heterogeneous} model has a higher score than the less elaborated \texttt{concat} model no matter what threshold was set. However, this measure lacks the interpretability of its value (Y-coordinate of curves on Figure \ref{fig:averaging_measures}).

\subsection{Ranking measures}\label{exp:ranking_measures}
\begin{figure}[t!]
\includegraphics[width=1.0\textwidth]{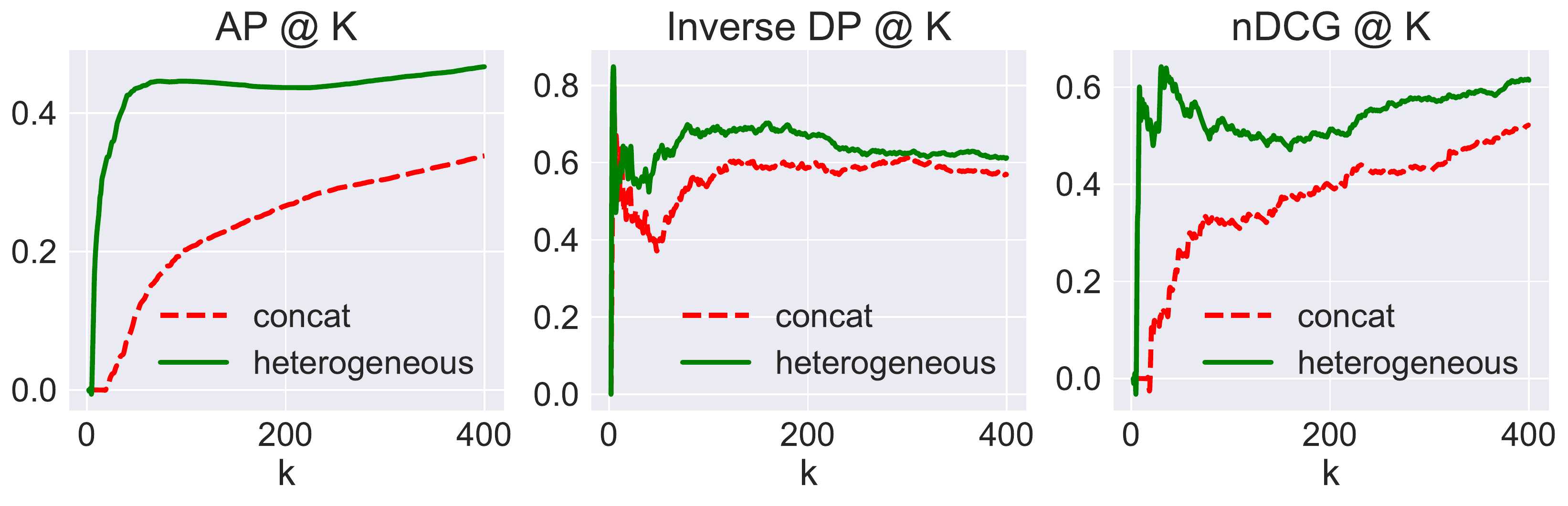}
\caption{\label{fig:ranking_stats}The ranking quality for the EmbedSim edge measure. The considered models (\texttt{concat} and \texttt{heterogeneous}) are described in Section \ref{algos}.}
\end{figure}

Figure \ref{fig:ranking_stats} shows that in all cases the ranking quality scores are, again, higher for \texttt{heterogeneous} model. One may interpret this result as the following: if a model is ``good'', than its top-$k$ edges should match the top-$k$ edges of the measure precisely enough, no matter what $k$ was set. According to Figure \ref{fig:ranking_stats} it holds for all ranking measures, but the biggest gap was given by the Average Precision. Hence, if one wants to compare quality of two different hierarchies, the advice may be the following: take Embedding similarity (EmbedSim) as the edge measure and plot the Average Precision @ $k$ graph. The better model will be the one having better score(s) at the desirable value(s) of $k$.

There is also a notable advantage of ranking approach over the averaging approach: it allows to choose the optimal number of hierarchical edges in the model, which we will discuss in the following subsection.

\subsection{Automated quality improvement}
\begin{figure}
\centering
\begin{subfigure}{.5\textwidth}
\centering
\includegraphics[width=1.0\textwidth]{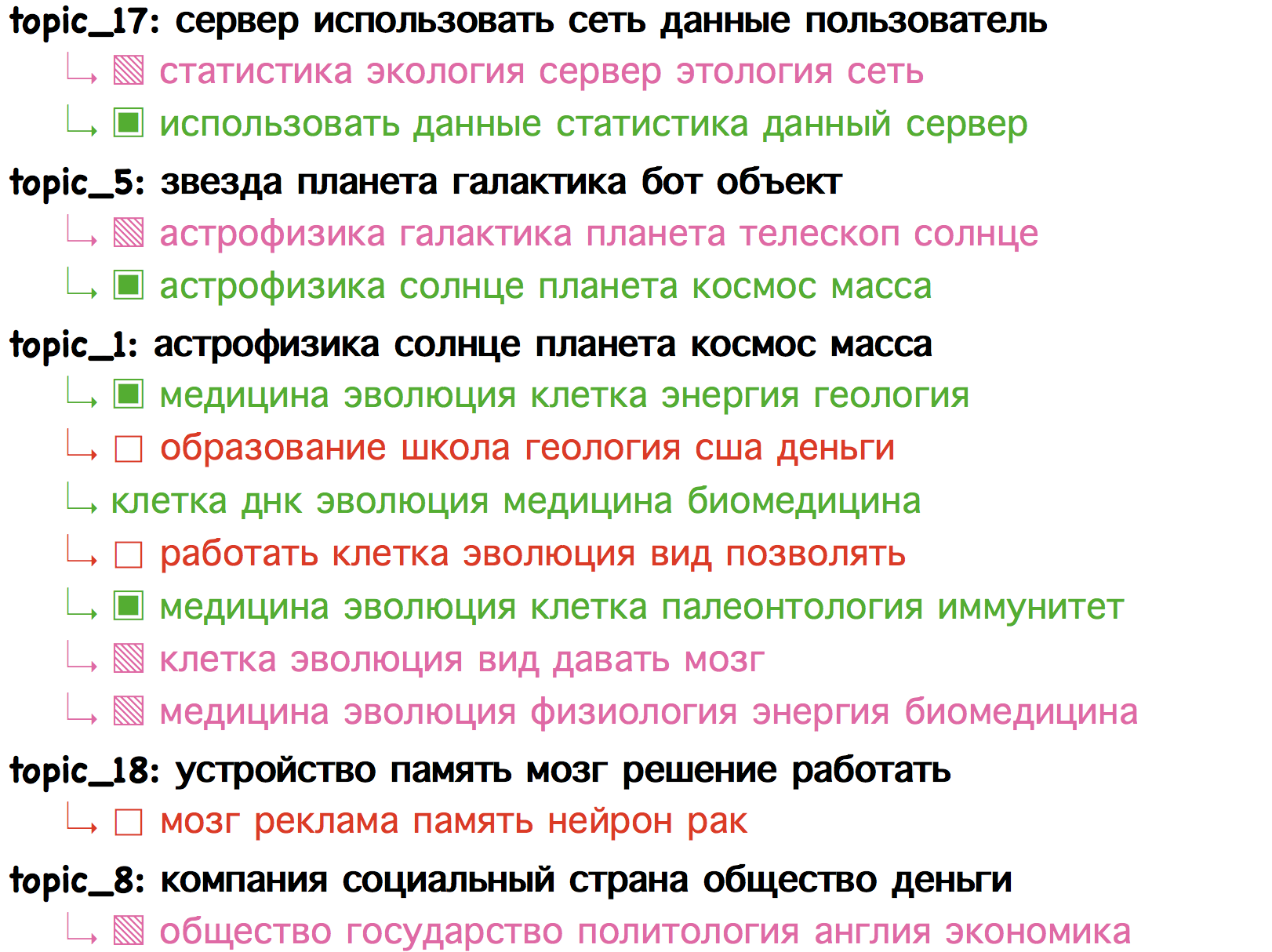}
\end{subfigure}%
\begin{subfigure}{.5\textwidth}
\centering
\includegraphics[width=1.0\textwidth]{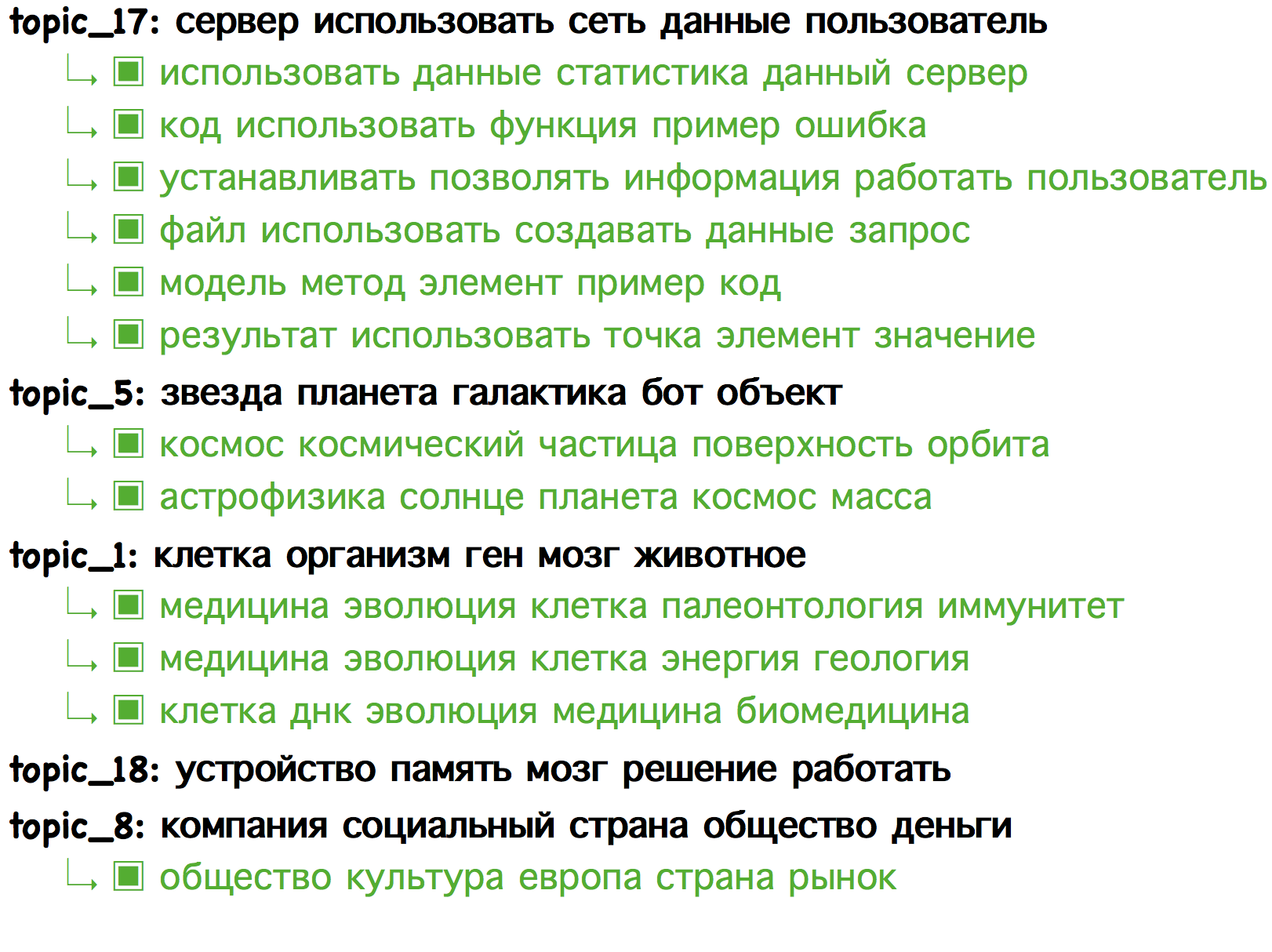}
\end{subfigure}
\caption{\label{fig:improvement}\textbf{(left)} A subset of first level (parent) topics of a hierarchy with their respective subtopics. Each topic or subtopic is represented by its 10 top tokens. The presented edges are divided into ``good'' (empty circles), ``bad'' (filled circles) and ``moderate'' (semi-filled circles). \textbf{(right)} The hierarchy from the left with the same first level (parent) topics presented. The new subtopics are chosen through ranking approach to edge selection.}
\end{figure}

The left side of Figure \ref{fig:improvement} demonstrates a subset of the parent topics of the  \texttt{concat} model with their child topics. According to our method, we plotted an inverse DP@$k$ graph for the EmbedSim measure of the edges of this model (see Section \ref{exp:ranking_measures}), found its maximum (in our case it was at $k = 100$) and built a new hierarchy that contained only the top-$k$ of the edges. The right side of Figure \ref{fig:improvement} demonstrates how quality of the same model increased without rebuilding the model itself. One can see that the new hierarchy looks more consistent and elaborated in comparison with the previous one.

\chapterconclusion
In Chapter 3, we conducted experiments on three Russian popular scientific datasets to validate that the measures, proposed in the previous chapter, are in agreement with human judgement on topical edges quality. We also compared the algorithms for heterogeneous topic modeling defined in Section \ref{algos} to find out that more elaborate \texttt{heterogeneous} algorithm outperforms the basic \texttt{concat} approach. Finally, we have presented an empirical method that improves the quality of already built models using ranking quality approach.

\startconclusionpage
In this work, we proposed several automated measures for ``parent-child'' relations of a topic hierarchy. We showed that the EmbedSim measure based on word embeddings reaches significant consistency with the assessors' judgment on whether the connection between topics exists or not. Other measures demonstrated moderate yet acceptable consistency and can also be used in conjunction with EmbedSim.

We also proposed two approaches for measuring quality of a hierarchy as a whole. Using measures of edges' quality we examined averaging and ranking approach to build an aggregated quality measure, and showed that better models reach higher scores in comparison with less elaborated models.

Finally, we demonstrated several applications of the proposed framework. First and foremost, we implemented an exploratory search engine \texttt{Rysearch}, which uses hierarchical topic modeling and the proposed algorithm to visualize a popular scientific topics in a hierarchical cell map. Another example is the usage of the proposed ranking approach for choosing the optimal set of edges to be included into a hierarchy.

Our work extends existing quality measures from flat topic models to hierarchical ones which, to the best of our knowledge, hasn't been done before. The results were presented at the \textit{24rd International Conference on Computational Linguistics and Intellectual Technologies (Dialogue'2018)} and at the \textit{60th Scientific Conference of MIPT}, winning the best paper award in the latter.

We can thus conclude that the research goal has been achieved and all research tasks have been accomplished.

\printmainbibliography

\end{document}